# Preserving data privacy when using multi-site data to estimate individualized treatment rules[1]

Coraline Danieli[a,*], Erica E. M. Moodie[a]

[a]McGill University, Department of Epidemiology, Biostatistics and Occupational Health

* coraline.danieli@rimuhc.ca

[1] **Funding:** This work was supported by the Canadian Institutes of Health Research grant CIHR TD3-137716 and the Natural Sciences and Engineering Research Council of Canada grant NSERC 228203. Erica E. M. Moodie is a Canada Research Chair (Tier 1) in Statistical Methods of Precision medicine. She acknowledges funding through a Discovery Grant from the Natural Science and Engineering Research Council of Canada (NSERC RGPIN-2019-04230) and salary support from a chercheur de mérite from the Fonds de recherche du Québec – Santé (FRQS-S CB Sr 34840).

**Abstract**

Precision medicine is a rapidly expanding area of health research wherein patient level information is used to inform treatment decisions. A statistical framework helps to formalize the individualization of treatment decisions that characterize personalized management plans. Numerous methods have been proposed to estimate individualized treatment rules that optimize expected patient outcomes, many of which have desirable properties such as robustness to model misspecification. However, while individual data are essential in this context, there may be concerns about data confidentiality, particularly in multi-centre studies where data are shared externally. To address this issue, we compared two approaches to privacy preservation: (i) data pooling, which is a covariate microaggregation technique and (ii) distributed regression. These approaches were combined with the doubly robust yet user-friendly method of dynamic weighted ordinary least squares to estimate individualized treatment rules. In simulations, we extensively evaluated the performance of the methods in estimating the parameters of the decision rule under different assumptions. The results demonstrate that double robustness is not maintained in data pooling setting and that this can result in bias, whereas the distributed regression provides good performance. We illustrate the methods via an analysis of optimal Warfarin dosing using data from the International Warfarin Consortium.

# 1. Introduction

Precision medicine is a rapidly expanding area of health research wherein patient level information is used to inform treatment decisions. Dynamic treatment regimens (DTRs) provide a statistical framework to formalize the individualization of treatment decisions that characterize personalized management plans. The term DTR is used to refer to treatment pathways or sequences that provide treatment recommendations over multiple treatment decision points dynamically, based on evolving patient characteristics; an individualized treatment rule (ITR) is treatment recommendation at a single time-point (e.g., at diagnosis) and is the focus of this work.

Numerous methods have been proposed to estimate DTRs that optimize expected patient outcomes [1-10], many of which have desirable properties such as robustness to model misspecification. However, while individual data are essential in this context, there may be concerns about data confidentiality, particularly in multi-centre studies (e.g., provincial electronic health record repositories, or hospitals in a multi-centre cohort study) where data are typically shared externally. As noted by Saha-Chaudhuri and Weinberg [11], the growth of personalized medicine has resulted in the collection of large quantities of very sensitive data for large segments of the population. Although these data may provide critical information for many aspects of health research, their use is constrained by legitimate concerns about privacy. Most standard data analysis techniques require individual data for estimation and inference, and thus are not applicable when individual data cannot be shared without risking disclosure of sensitive information. Frequently, data for research undergo anonymization to remove identifying information. However, one may still be able to identify participants using combinations of variables or auxiliary information available from other sources [12-13]. Sometimes data are deliberately coarsened to obscure individual identities, leading to loss of precision and statistical power. A key challenge of data privacy preservation is to strike a good balance between data usability and confidentiality [11].

Different strategies have been proposed that aim to protect the confidentiality of the individual data while disseminating useful information derived from these data, mostly in traditional regression analyses. Data manipulations such as top-coding, cell suppression, data swapping, and jittering are all different techniques that can help to avoid identity or attribute disclosures [14]. Other strategies use transformation of the data in order to hide sensitive information. These strategies include the randomization method [15], which adds a noise in the data in such a manner that each centre can learn only the minimum possible about the other centres. This method is for instance used in the secure summation protocol

[16], and microaggregation [17], which represents a family of techniques that proposes to replace individual values by values computed on small aggregates. Several methods stem from this basic concept of aggregation such as basic data poolings, k-anonymity, or l-diversity. These methods differ in the way that they minimize the information loss caused by microaggregation, which depends on the level of confidentiality required and on the tolerance for loss of statistical power.

There have been also ongoing efforts to develop methods that enable researchers to conduct multi-database regression analysis without the need to centrally combine all individual-level data from participating sites [14,18]. Among these methods is the distributed regression approach [14,19]. Distributed regression performs the same numeric algorithm as standard regression that is based on individual-level data but uses only summary matrix products (obtained with few manipulations from each centre for computation. This approach, therefore, typically involves a high level of data aggregation and the sharing of a smaller number of data summaries while again avoiding sharing of individual-level data. For linear models, distributed regression and individual-level data analysis should produce statistically equivalent results.

Although individualized treatment strategies can be identified via regressions with interactions between the treatment and the covariates, concerns about data privacy have never been explored in this setting. Therefore, to address data privacy in the context of estimating individualized treatment rules from multi-site studies, we focused on two approaches: (i) data pooling and (ii) distributed regression, where we adapt both to a doubly-robust ITR estimation known as Generalized dynamic Weighted Ordinary Least Squares (G-dWOLS) [8,20]. G-dWOLS is an approach which bears similarities to both Q-learning [1-2], affording simplicity of implementation, intuitiveness, and many model-diagnostic tools [21], and the more complex and doubly robust G-estimation [4].

In this paper, we adapt and compare the afore-mentioned two approaches to address concerns surrounding data sharing when attempting to use data from multiple centres to estimate optimal individualized treatment rules in the G-dWOLS framework. After briefly summarizing the G-dWOLS approach, Section 2 describes how data pooling and distributed regression can be used to maintain privacy preservation in an ITR context. Section 3 demonstrates the performance of the methods via a series of simulations under a wide variety of different scenarios. Section 4 applies both approaches to estimate an optimal individualized Warfarin dose strategy in the context of data confidentiality. We conclude the paper with a discussion.

## 2. Methods

### A. The Generalized dynamic Weighted Ordinary Least Squares approach

Generalized dynamic Weighted Ordinary Least Squares [20] is an extension to the continuous treatment setting of the dynamic weighted ordinary least squares approach[8] (dWOLS), one of the main methods for regression-based DTR estimation originally developed only for a binary treatment context. In the multiple decision-point setting, G-dWOLS sequentially determines the optimal treatment decision across treatment stages and indirectly allows for the identification of the optimal DTR using a backwards induction approach. Here, only a single-stage setting will be considered. This greatly simplifies the question of interest as well as the G-dWOLS approach, as it reduces to a single, weighted least squares estimation procedure.

Consistent with the DTR literature, let $Y$ denote a continuous outcome, where larger values of $Y$ are clinically preferable. The assigned treatment will be denoted as $A$, whether binary or continuous. Subject-specific (or individual history) covariates will be denoted by the random vector $X$.

The aim is to find the optimal treatment $A^{opt}$ that maximizes the expected outcome. We conceptualize the outcome model, $E(Y|X, A)$, as being split in two components: the "treatment-free" function $f$, and the "blip" function $\gamma$. This may be written as:

$$E(Y|X = x, A = a) = f(x^\beta, \beta) + \gamma(x^\psi, a; \psi) \qquad (1)$$

where $x^\beta$ and $x^\psi$ are subsets of the observed vector $x$ (and $x^\psi$ is contained in $x^\beta$), the "treatment-free" function $f$ represents the impact, in absence of treatment (or at the reference level of treatment), of patient history $x^\beta$ on the outcome (for instance, when linear, $f(x^\beta, \beta) = \beta_0 + \beta_1 x^\beta$), and the "blip" function represents the impact of the treatment on the outcome, including both the main effect and the interaction between the treatment and the covariates $x^\psi$. The blip, or contrast, function must be parameterized so as to equal zero when $A=0$, whether treatment is binary or otherwise. With a known or estimated blip function, it is straightforward to determine the optimal treatment rule as that which maximizes the blip function, since a maximized blip function will simultaneously maximize the conditional expectation of the outcome. Thus, estimation of the blip function parameters is synonymous with estimation of the ITR itself.

Therefore, since this blip function represents the only means by which the treatment can influence the expected outcome, we need only to focus on the form and parameters of $\gamma$

(for instance, when linear, $\gamma(x^\psi, a; \psi) = a(\psi_0 + \psi_1 x^\psi)$, estimation of $\psi_0$ and $\psi_1$ is of primary interest). However, since the treatment may depend on the patient history, a third function, called the treatment density function or (generalized) propensity score, $f[A|X; \alpha]$, needs to be defined in the G-dWOLS framework.

G-dWOLS relies on several standard assumptions: no interference between subjects [22], no unmeasured confounding [23], positivity [3], and lack of other biases in the data including no measurement error or selection bias in the sampled units. Any missing data are, for the purposes of our development, assumed to be missing completely at random. Assuming there are no unmeasured confounders, G-dWOLS eliminates bias due to confounding via either a correctly-specified treatment-free model or through weighting based on a function of the propensity score. Accordingly, G-estimation allows for doubly robust estimation of the blip model parameters as only one (or both) of either the treatment-free or treatment models needs to be correctly specified for consistent estimation of the blip model parameters.However, the blip function must be correctly specified.

G-dWOLS achieves consistent estimation of the blip model parameters by employing a weighted linear regression to fit the model in Equation (1) with weights $w$ derived from a correctly identified (generalized) propensity score model [24] satisfying specific conditions [20]. In binary treatment setting, the weight $w = |A - E[A|X; \hat{\alpha}]|$ has been shown to work well; the well-known inverse probability of treatment weights also satisfy the balancing conditions required by G-dWOLS. In continuous treatment setting, a larger range of weight functions can be applied [20].

## B. Data pooling

In multi-centre studies where data are shared externally, data confidentiality may be a critical issue. To address this problem, microaggregation techniques allow aggregating the covariates values within groups of a specific centre or over centres so that the information from all centres can be used for the statistical analyses without sharing individual-level data. Aggregating data often consists of summing individual-level covariates. The choice of the group size depends on the level of confidentiality required and on the tolerance for loss of statistical power: when smaller group size is used, the level of the confidentiality may not be optimal but the statistical power may be higher [11]. In the following, we will focus on three different data pooling strategies [11], which are demonstrated by a small example in Table 1.

The *first data pooling strategy* proposes the formation of pools (or groups) of size $g$ across the centres, and the aggregation (that is the sum) of the covariates within each pool.

One limitation of this method is that it requires collaboration between centres to create the pools (hence some data sharing between centres) to reach the analyses stage and thus would require additional confidentiality measures to ensure privacy. The *second data pooling strategy* proposes that each centre randomly creates their own pools of size $g$ ($g$ being common for each centre) and then aggregates the covariates within each pool. One limitation of this method is that if there is no common divisor between centres, some patients could be excluded and thus precision (power) may be lost due to unused data, in addition to any loss of power due to the aggregation itself. The *third data pooling strategy* proposes that each centre randomly creates their own pools of size $g_k$ (where $g_k$ is allowed to differ between centres) and then aggregates the covariates within each pool. This method would reduce the number of patients excluded, as it is the case for the second strategy.

**Table 1.** Schema of the different pooling strategies with three centres and 25 patients distributed over these centres. The third column indicates the pool number when the first pooling strategy is applied: each patient is randomly assigned to a specific pool of size $g = 5$. The fourth column indicates the pool number when the second pooling strategy is applied: within their own centre, each patient is randomly assigned to a specific pool of size $g = 3$ (common for all centres). Since the first centre has four patients, one person is excluded (patient 3) and his data will not be used for the analyses. The fifth column indicates the pool number when the third pooling strategy is applied: within their own centre, each patient is randomly assigned to a specific pool of size $g_1 = 2$ for centre 1, $g_2 = 3$ for centre 2, and $g_3 = 5$ for centre 3. Since every centre can decide their own pool size, there are fewer or no exclusions of patient data used in analyses.

| Patient | Centre | Pool First strategy ($g = 5$) | Pool Second strategy ($g = 3$) | Pool Third strategy ($g_1 = 2, g_2 = 3, g_3 = 5$) |
|---|---|---|---|---|
| 1 | 1 | 1 | 1 | 1 |
| 2 | 1 | 3 | 1 | 2 |
| 3 | 1 | 2 | - | 2 |
| 4 | 1 | 2 | 1 | 1 |
| 5 | 2 | 3 | 2 | 3 |
| 6 | 2 | 5 | 3 | 3 |
| 7 | 2 | 1 | 2 | 3 |
| 8 | 2 | 5 | 2 | 4 |
| 9 | 2 | 5 | 3 | 4 |
| 10 | 2 | 4 | 3 | 4 |
| 11 | 3 | 4 | 4 | 5 |
| 12 | 3 | 1 | 6 | 6 |
| 13 | 3 | 3 | 7 | 7 |
| 14 | 3 | 4 | 6 | 7 |
| 15 | 3 | 2 | 8 | 6 |

| 16 | 3 | 4 | 4 | 7 |
| --- | --- | --- | --- | --- |
| 17 | 3 | 5 | 8 | 6 |
| 18 | 3 | 1 | 8 | 5 |
| 19 | 3 | 5 | 5 | 7 |
| 20 | 3 | 3 | 7 | 6 |
| 21 | 3 | 4 | 7 | 5 |
| 22 | 3 | 2 | 5 | 7 |
| 23 | 3 | 3 | 4 | 5 |
| 24 | 3 | 2 | 6 | 5 |
| 25 | 3 | 1 | 5 | 6 |

Saha-Chaudhuri and Weinberg [11] applied these data pooling approaches to analyze confidential data, particularly in the context of a matched case-control study [26-27], with a view to estimating the individual-level odds ratios of an exposure of interest using a conditional logistic regression model. The simulation study showed encouraging results that were similar between pooled and unpooled analyses, whatever the pools size, in terms of bias, standard errors and coverage probability [11].

In our case, we intend to use this approach to estimate the individual-level blip parameters, and hence the ITR. To illustrate our goal, we consider the following outcome model, consisted of a linear treatment-free function $f(x^\beta, \beta) = \beta_0 + \beta_1 x^\beta$, and a linear blip function $\gamma(x^\psi, a; \psi) = a(\psi_0 + \psi_1 x^\psi)$.

$$E(Y|X = x, A = a) = (\beta_0 + \beta_1 x^\beta) + a(\psi_0 + \psi_1 x^\psi) \qquad (2)$$

Each component of the model will be aggregated within a particular pool of size $g$ for instance, defined by following one of the pooling strategies presented above. The outcome model would become then:

$$\sum_{i=1}^{g} E(Y_i|X_i = x_i, A_i = a_i) = \sum_{i=1}^{g}(\beta_0 + \beta_1 x_i^\beta) + \sum_{i=1}^{g} a_i(\psi_0 + \psi_1 x_i^\psi)$$

$$= \sum_{i=1}^{g} \beta_0 + \beta_1 \sum_{i=1}^{g} x_i^\beta + \psi_0 \sum_{i=1}^{g} a_i + \psi_1 \sum_{i=1}^{g} a_i x_i^\psi$$

$$= \sum_{i=1}^{g} \beta_0 + \beta_1 \sum_{i=1}^{g} x_i^\beta + \sum_{i=1}^{g} a_i \left( \psi_0 + \psi_1 \frac{\sum_{i=1}^{g} a_i x_i^\psi}{\sum_{i=1}^{g} a_i} \right)$$

$$E(Y_{pool}|X_{pool} = x_{pool}, A_{pool} = a_{pool}) = \beta_{0,pool} + \beta_1 x^{\beta}_{pool} \quad (3)$$
$$+ a_{pool}(\psi_0 + \psi_1 x^{\psi^w}_{pool})$$

with $\beta_{0,pool} = \sum_{i=1}^{g} \beta_0$, $x^{\beta}_{pool} = \sum_{i=1}^{g} x_i^{\beta}$, $a_{pool} = \sum_{i=1}^{g} a_i$ and $x^{\psi^w}_{pool} = \sum_{i=1}^{g} a_i x_i^{\psi} / \sum_{i=1}^{g} a_i$ (a weighted version of $x^{\psi}$). In the "pooled" outcome model defined in Equation (3), the parameters of interest coincide with the parameters of the "pooled" blip function, $\gamma(x^{\psi^w}_{pool}, a_{pool}; \psi) = a_{pool}(\psi_0 + \psi_1 x^{\psi^w}_{pool})$, and these correspond to the parameters of the initial outcome model in Equation (2). Therefore, using data pooling strategy would allow data confidentiality while estimating the individual-level blip parameters.

## C. Distributed regression

We now describe the distributed regression approach to preserving privacy while sharing data appropriate for G-dWOLS estimation of an ITR. For simplicity, consider three centres $m = \{1,2,3\}$ and suppose we are, as before, interested in the outcome model – and most particularly in the blip function parameters – of Equation (2) for each centre.

When each centre does their analyses separately, at the individual-data level, using a matrix formulation, we express these as:

$$y_1 = X_1^{\beta} \beta_{(1)} + X_1^{\psi} \psi_{(1)} + \varepsilon_1 = X_1 \theta_{(1)} + \varepsilon_1$$
$$y_2 = X_2^{\beta} \beta_{(2)} + X_1^{\psi} \psi_{(2)} + \varepsilon_2 = X_2 \theta_{(2)} + \varepsilon_2$$
$$y_3 = X_3^{\beta} \beta_{(3)} + X_3^{\psi} \psi_{(3)} + \varepsilon_3 = X_3 \theta_{(3)} + \varepsilon_3$$

where $y_m$ represents the vector of outcomes for centre $m$, $X_m^{\beta}$ represents the design matrix related to the treatment-free function for centre $m$, $\beta_{(m)}$ represents the vector of parameters associated with $X_m^{\beta}$, $X_m^{\psi}$ represents the design matrix related to the blip function for centre $m$ (with $A_m$ already incorporated into $X_m^{\psi}$), $\psi_{(m)}$ represents the vector of parameters associated with $X_m^{\psi}$, $X_m$ represents the combination of $X_m^{\beta}$ and $X_m^{\psi}$, with its vector of parameters $\theta_{(m)} = (\beta_{(m)}, \psi_{(m)})$ and $\varepsilon_m$ represents the random error for centre $m$. The usual weighted least square estimation approach would estimate the parameters of interest, that is, $\boldsymbol{\theta}_{(1)}$, $\boldsymbol{\theta}_{(2)}$, and $\boldsymbol{\theta}_{(3)}$, for each centre separately in the context of data privacy. Estimates of these parameters could then potentially be combined via meta-analytic techniques without needing to share individual-level data across centres. However, it has been shown that one of the limitations of traditional meta-analysis methods arises for detection of subgroup association under heterogeneity [28-30]. Therefore, it is unclear that this approach is suitable for ITRs, where heterogeneity in treatment response is particularly of interest and meta-analyses may not be able to detect such effects [31-32].

Dealing with data privacy in a different manner, the G-dWOLS with distributed regression approach targets the parameters of interest directly [14,19]:

$$\begin{pmatrix} y_1 \\ y_2 \\ y_3 \end{pmatrix} = \begin{pmatrix} X_1 \\ X_2 \\ X_3 \end{pmatrix} \times \boldsymbol{\theta} + \begin{pmatrix} \varepsilon_1 \\ \varepsilon_2 \\ \varepsilon_3 \end{pmatrix}.$$

Obtaining estimates of such parameters with individual-level data is straightforward. However, in the context of privacy preservation, obtaining G-dWOLS estimates with the distributed regression approach requires minimizing the sum over the centres of the error sum of squares:

$$S = \sum_{i=1}^{n_1} w_{1i}\varepsilon_1^2 + \sum_{i=1}^{n_2} w_{2i}\varepsilon_2^2 + \sum_{i=1}^{n_3} w_{3i}\varepsilon_3^2,$$

which we can re-express as

$$S = W_1[y_1 - X_1\boldsymbol{\theta}]'[y_1 - X_1\boldsymbol{\theta}] + W_2[y_2 - X_2\boldsymbol{\theta}]'[y_2 - X_2\boldsymbol{\theta}] + W_3[y_3 - X_3\boldsymbol{\theta}]'[y_3 - X_3\boldsymbol{\theta}]$$

$$S = [y_1 W_1 y_1' + y_2 W_2 y_2' + y_3 W_3 y_3'] - 2[X_1' W_1 y_1 + X_2' W_2 y_2 + X_3' W_3 y_3]\boldsymbol{\theta} + \boldsymbol{\theta}[X_1' W_1 X_1 + X_2' W_2 X_2 + X_3' W_3 X_3]\boldsymbol{\theta}$$

where $w_{mi}$ corresponds to the weight of $y_{mi}$, $m = \{1,2,3\}$. Using the first derivative of **S** over $\boldsymbol{\theta}$ to obtain the extremum gives

$$\frac{\partial S}{\partial \boldsymbol{\theta}} = 0 \Leftrightarrow [X_1' W_1 y_1 + X_2' W_2 y_2 + X_3' W_3 y_3] = \boldsymbol{\theta}[X_1' W_1 X_1 + X_2' W_2 X_2 + X_3' W_3 X_3],$$

and so the weighted least squares estimator $\widehat{\boldsymbol{\psi}}$ can be derived from the weighted least squares estimator $\widehat{\boldsymbol{\theta}} = (\widehat{\boldsymbol{\beta}}, \widehat{\boldsymbol{\psi}})$, which is given by

$$\widehat{\boldsymbol{\theta}} = [X_1'W_1X_1 + X_2'W_2X_2 + X_3'W_3X_3]^{-1}[X_1'W_1y_1 + X_2'W_2y_2 + X_3'W_3y_3].$$

Therefore, to estimate these treatment-decision parameters, each centre $m$ must provide two matrix products: $X_m'W_mX_m$ and $X_m'W_my_m$, which would allow centres to keep all individual-level data confidential while still sharing information from all site participants. In each centre, this approach requires only estimation of the weight matrix $W_m$ (which itself requires estimation of the propensity score) and very little analytic sophistication beyond what is required for, say, data pooling. Note that in the distributed regression approach, the weights are assigned at the individual level and then aggregated into the matrix to be transmitted to the central analysis site. This is contrary to the data pooling strategies, for which the weights are assigned at the pool-level, after the data aggregation step.

## 3. Simulation studies

The simulation study is reported following the ADEMP (aims, data-generating mechanisms, estimands, methods, and performance measures) scheme proposed by Morris, White and Crowther [33].

### *A. Aims*

Our primary objective is to adapt ITR estimation to address data privacy concerns. Accordingly, our simulations evaluated the performance of G-dWOLS for estimation of treatment decision rule parameters using the two proposed approaches to address data privacy, under different assumptions concerning (1) the distribution of the subject-specific covariate in each centre, (2) the treatment variable distribution, (3) the degree of confounding, captured by the strength of relationship between the treatment variable and the subject-specific variable, (4) the pool size when aggregating the data and (5) the model misspecification. Points (1)-(4) concerns data-generating mechanisms, while (5) concerns the analyses.

*B. Data-generating mechanisms*

We considered multiple data-generating mechanisms. For each, we considered a single-stage (ITR rather than DTR) setting. The multiple data-generating mechanisms represented different combinations of:
1. The similarity or the difference of the distribution of the subject-specific covariate $X$ in each centre
    a. For all centres, the subject-specific covariate $X$ follows a normal distribution N(10,1); this is a *homogeneity assumption* across sites.
    b. The subject-specific covariate $X$ follows a N(10,1) distribution, a uniform law U[6,14] or a log-normal law LN(0.7,0.5) according to the centre to which they belong. This setting allows for the more realistic possibility that covariate distributions vary across centres while still assuming broadly comparability (similar means and ranges) across sites so that the populations are "overlapping" though not identical.
2. The treatment variable distribution
    a. Binary treatment: $A \sim B(n, p(x))$, where $p(x)$ is as given in 3, below.
    b. Continuous treatment: $A \sim N(X, SD_A)$.
3. The degree of confounding, as captured by the strength of relationship between $A$ and $X$.
    a. The probability of being treated when the treatment is binary, which was set by the coefficient $\rho$ of the following function that represents the probability of being treated according to $X$, $p(x) = 1/(1 + \rho e^{-(x-10)})$:
        i. "Lowest level": $\rho = 15$ (yielding 5-20% of individuals treated)
        ii. "Low level": $\rho = 8$ (yielding 10%-25% of individuals treated)
        iii. "Moderate level": $\rho = 5$ (yielding 15%-30% of individuals treated)
        iv. "High level": $\rho = 2$ (yielding 25%-40% of individuals treated)
        v. "Highest level": $\rho = 1$ (yielding 35%-50% of individuals treated)
    b. When the treatment is continuous, the level of the confounding was specified through the correlation between $A$ and $X$, which in turn was set by the standard deviation of the treatment variable $SD_A$:
        i. Very low correlation: $SD_A = 12$ which corresponds to $0 \leq cor(A,X) \leq 0.1$
        ii. Low correlation: $SD_A = 4$ which corresponds to $0.2 \leq cor(A,X) \leq 0.3$
        iii. Moderate: $SD_A = 1.9$ which corresponds to $0.4 \leq cor(A,X) \leq 0.5$
        iv. High: $SD_A = 1.2$ which corresponds to $0.6 \leq cor(A,X) \leq 0.7$

v. Very high: $SD_A = 0.5$ which corresponds to $0.8 \leq cor(A,X) \leq 1$.

The outcome model was taken to be:

$$Y_i = log(X_i) + sin(X_i) + X_i + A_i(\psi_0 + \psi_1 X_i) + \varepsilon_i,$$

where $log(X_i) + sin(X_i) + X_i$ represents the treatment-free function, $A_i(\psi_0 + \psi_1 X_i)$ represented the blip function with $\psi_0 = \psi_1 = 1$, and the random errors $\varepsilon_i$ follow a standard normal law, N(0,1).

Using the data pooling strategy, after having decided upon a pool size $g$, the $N/g$ pools were formed by randomly partitioning the $N$ patients into pools of size $g$ [11], assuming $N$ is a multiple of $g$. (Groups are usually formed using a criterion of maximal similarities, where the similarities can be measured using a distance. The optimal partition [17] is defined to be the one that maximizes the within-group homogeneity. The higher the within-group homogeneity, the lower the information loss, where the sum of squares criterion is commonly used to measure homogeneity in clustering.) When pooling the data, we formed pooled covariates, where each one corresponded to the sum of the specific original covariate values over the $g$ patients. The pooled outcome model was then expressed as:

$$y_{pool} \sim logx_{pool} + sinx_{pool} + x_{pool} + a_{pool}(\psi_0 + \psi_1 x^w_{pool}) + \varepsilon_{pool},$$

with $y_{pool} = \sum_{i=1}^{g} y_i$, $logx_{pool} = \sum_{i=1}^{g} log(x_i)$, $sinx_{pool} = \sum_{i=1}^{g} sin(x_i)$, $x_{pool} = \sum_{i=1}^{g} x_i$, $a_{pool} = \sum_{i=1}^{g} a_i$, $x^w_{pool} = \sum_{i=1}^{n_k} a_i x_i / \sum_{i=1}^{n_k} a_i$ and $\varepsilon_{pool} = \sum_{i=1}^{g} \varepsilon_i$.

For each of the 40 simulated scenarios summarized in Table S1 in Section 1 of the Online Supplementary Material (scenarios a-t, a.bis-j.bis, a.ter-j.ter), we generated 1000 independent data sets of size $N = 60,000$, divided in three centres. Using the data pooling strategy, original data were then aggregated into $N/g = 2000/600/100$ pools of size $g = 30/100/600$ patients respectively. All simulations were performed using a customized program in R software that is available in Section 2 of the Online Supplementary Material.

*C. Estimands, methods, and performance metrics*

The estimands of interest were the blip function parameters $\psi_0$ and $\psi_1$ as the blip function is the only means by which treatment can influence the expected outcome, and hence fully characterizes the optimal treatment strategy.

All analyses relied on the first data pooling strategy and on the distributed regression approach, each combined with G-dWOLS using the inverse probability of treatment weight for continuous treatment setting or with $|a - E[A|X]|$ as weight for binary treatment setting. (Note that in the binary treatment setting where individuals' treatments consist of values 0 (untreated) or 1 (treated), the treatment in the pooled sample follows a binomial distribution B($n,p$) (with parameters $n$ = pool size and $p$ = number of patients within the pool that are assigned a treatment). Therefore, a binomial generalized linear model (GLM) adjusting for the aggregated covariates of interest (sum of the values of the covariates over the patients constituting the pool) is used to be estimate the treatment model. The weights are then estimated as the difference between the observed treatment in the pooled sample and the predicted values obtained from the GLM.

In the continuous treatment setting where individuals' treatments consist of continuous-valued doses, the treatment in the pooled sample follows a normal distribution. Therefore, a linear model adjusting for the aggregated covariates of interest (sum of the values of the covariates over the patients constituting the pool) is the relevant approach to use as the aggregated data generalized propensity score model. We choose to estimate the weights using the inverse probability of treatment from the predicted values obtained from the linear model. Those weights are then used in a weighted linear regression of the outcome $y_{pool}$ on $logx_{pool}, sinx_{pool}, x_{pool}, a_{pool}$, and $a_{pool}x_{pool}^w$.

The performance of G-dWOLS using the original (individual-level) data, data pooling, and distributed regression was assessed under various forms of model misspecification (always assuming a correctly-specified blip model):
- Scenario 1. Both the treatment and treatment-free models are correctly specified
    - Treatment model: $logit(p(1|x)) = x$ if the treatment is binary and $A = x$ if the treatment is continuous
    - Treatment-free model: $f(x) = log(x) + sin(x) + x$
- Scenario 2. The treatment model is correctly specified but the treatment-free model is misspecified
    - Treatment model: $logit(p(1|x)) = x$ if the treatment is binary and $A = x$ if the treatment is continuous
    - Treatment-free model: $f(x) = x$

- Scenario 3. The treatment model is misspecified but the treatment-free model is correctly specified
    - Treatment model: $logit(p(1|x)) = 1$ if the treatment is binary and $A = \beta_0$ if the treatment is continuous
    - Treatment-free model: $f(x) = log(x) + sin(x) + x$
- Scenario 4. Both the treatment and treatment-free models are misspecified.
    - Treatment model: $logit(p(1|x)) = 1$ if the treatment is binary and $A = \beta_0$ if the treatment is continuous
    - Treatment-free model: $f(x) = x$

We assessed (i) the relative bias which represents the difference between the mean of the estimates and the corresponding true value, divided by the latter, and (ii) the empirical standard deviation. The results were compared with results obtained from the "gold standard" analysis using the original individual-level data.

## *D. Simulation results*

In this section, we present the results for ten selected simulated scenarios (a-j, see Table S1 for details). The results focus on the parameter $\psi_0$, which is more affected by the different studied scenarios than $\psi_1$, as the relative bias and the standard deviation are typically higher. The remaining results are presented in the Online Supplementary Materials.

For binary treatment distribution (Figure 1 and Table S2 available in Section 3 of the Online Supplementary Material), in both scenarios 1 and 3 where the treatment-free model is correctly specified, data pooling and distributed regression provide consistent estimation of the blip parameter $\psi_0$, with relative bias less than 3% whatever the level of confounding. Distributed regression performs similarly to individual-level data, whereas variability is a bit higher for data pooling. In scenario 2, where the treatment model is correctly specified, the double robustness of the G-dWOLS method is highlighted. As seen in Figure 1 and Table S2, data aggregation is again similar to estimation using individual-level data with relative bias less than 1% and variability similar to that obtained in the other scenarios. However, data pooling yields a biased estimator of the blip parameter $\psi_0$ (Figure 1 and Table S2), suggesting that the double robustness of G-dWOLS cannot be assured with pooled data. In scenario 4, where both the treatment and treatment-free models are misspecified, all methods yield biased estimator, with a substantial relative bias and similar standard deviations over the different scenarios (Figure 1 and Table S2). Across all scenarios, data pooling is far less efficient than distributed regression, as effective degrees of freedom are lost due to the data pooling.

For the continuous treatment ("dose") distribution (Figure 2 and Table S3 available in section 3 of the Online Supplementary Material), results are similar. Again, in scenarios 1 and 3 where the treatment-free model is correctly specified, there is very little relative bias (<2%), with some loss of precision seen with data pooling. In scenario 2, where the treatment model is correctly specified, consistent estimate of the blip parameter is observed when there is little confounding (very low correlation), however as the degree of confounding increases, relative bias exceeds 14% (Figure 2 and Table S3). This is likely due to higher correlations leading to large weights and unstable estimation. Across all scenarios, the standard errors varying according to the level of the confounding: the higher the correlation between the covariate and the treatment, the higher the variability of the estimates (Figure 2 and Table S3). Unsurprisingly, results are poor for all approaches when both models are misspecified (scenario 4). The results obtained from data pooling data when the pool size is increased to 100 or 600 exhibit a similar pattern to that observed when the pool size was 30 (results not shown).

The effect of varying the distribution of the subject-specific covariate over centres was also investigated (see Section 4 of the Online Supplementary Material). Whether treatment is binary (Table S4, Figure S1) or continuous (Table S5 and Figure S2), the results were broadly similar to those reported above, where the distribution of the subject-specific covariate does not vary across centres. The exceptions are that the data pooling performed rather more poorly than both the individual-level data and distributed regression in the binary data setting, and, in the continuous data setting with scenario 3 and a high degree of confounding, considerable instability in estimation with distributed regression was observed.

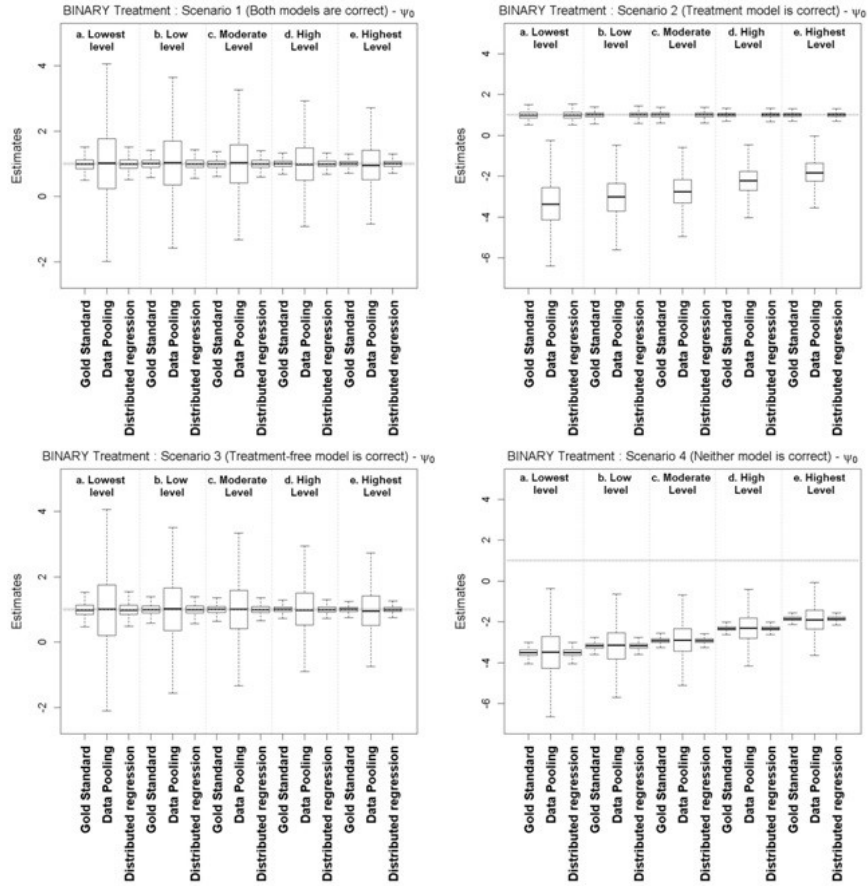

**Figure 1.** Simulation results: Performance of the methods over 1000 samples for each scenario, according to different degrees of confounding, when the treatment distribution is BINARY and the subject-specific covariates distribution is IDENTICAL. (The pool size when aggregating the data is 30.)

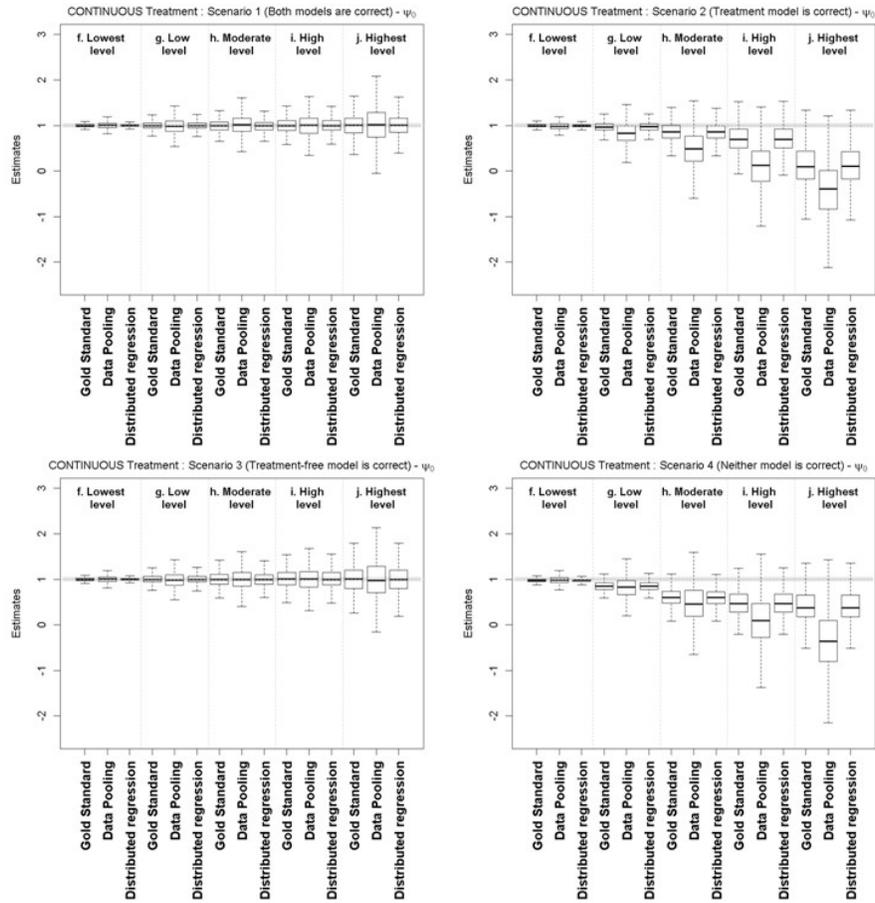

**Figure 2.** Simulation results: Performance of the methods over 1000 samples for each scenario, according to different degrees of confounding, when the treatment distribution is CONTINUOUS and the subject-specific covariates distribution is IDENTICAL. (The pool size when aggregating the data is 30.)

## 4. Application: Estimating an optimal Warfarin dose strategy

*A. Data*

Warfarin is a widely prescribed anticoagulant used to treat and prevent thrombosis and thromboembolism. Establishing a correct therapeutic dose is difficult as its effects are highly variable amongst patients.[34] In determining the ideal subject-specific Warfarin dose, a patient's international normalized ratio (INR) is closely monitored and the dose is varied until a stable therapeutic INR state is achieved, typically in the range of [2;3] [35-37]. INR is a measure of the time needed for blood to clot. If INR is high, this indicates that blood is clots more slowly which could lead to dangerous bleeding, whereas a low INR indicates clotting that is too fast which can have serious consequences such as thrombosis or a pulmonary embolism. The goal of The International Warfarin Pharmacogenetics Consortium [34] was to explore several models for determining an appropriate Warfarin dose algorithm. The researchers found that a pharmacogenetic model performed best, suggesting that the prescription of Warfarin should be tailored according to individual patient level genetic and clinical characteristics. Subsequent studies applied other methods and came to similar conclusions [38,39].

We revisited The International Warfarin Pharmacogenetics Consortium data [34] (openly available at [www.pharmgkb.org/page/iwpc](www.pharmgkb.org/page/iwpc)), gathered data from multiple centres, to apply the two proposed approaches to estimate an optimal individualized Warfarin treatment strategy in the G-dWOLS framework with a single stage setting in the context of data confidentiality. The results were compared with those obtained from the gold standard analysis using the original individual-level data. Following the work of Chen, Zeng, and Kosorok [38], Wallace, Moodie and Stephens [39], and Schulz and Moodie [20], observations with missing values were removed, leading to a sample size of n = 1732, originated from 11 different centres. However, two of these eleven centres were removed from the total cohort since they each included 1 and 4 patients respectively, which render data pooling infeasible in these centres. Therefore, the sample size was reduced to 1727. The final dataset used in the analysis included several predictors; specifically, the patient's age (binned into 9 groups), the weight centreed by the site mean, the height centreed by the site mean, an indicator for taking either Rifampin, Carbamazepine or Phenytoin (coded as enzyme = 1), an indicator for taking Amiodarone, gender ("Man" being the reference level), race (labelled as "Black" or "Asian" or "White", with the latter the reference level), and genetic information on the VKORC1 (3 levels) and CYP2C9 (4 levels) genotypes. These potential confounders and/or tailoring variables will be collectively denoted by the matrix $x$. Information on each patient's stable Warfarin dose and corresponding INR was also included in the data.

## B. Analyses

The analyses using the G-dWOLS method relied on a blip function that is quadratic in the Warfarin dose, including the main effects of both dose and squared-dose as well as their respective interactions with all of the aforementioned predictors ($x^\psi = x$). Thus, in accordance with previous work [20], the outcome model was specified as:

$$y = x\beta + a(\psi_{01} + x\psi_{11} + a\psi_{02} + ax\psi_{12}), \qquad (4)$$

where the outcome variable $y$ was defined as $y = -\sqrt{|2.5 - INR|}$, so that larger values of $y$ were clinically preferable (closer to zero, representing the mid-point of the therapeutic range of INR). The observed distribution of the outcomes was roughly symmetric, with values varying from -1.0954 to 0. The vector $\beta$ represented the effect of all predictors $x$ that specified the treatment-free function ($x^\beta = x$), on the outcome $y$. The vector $a$ represented the vector of dose, and $\psi_{01}$, $\psi_{02}$, $\psi_{11}$ and $\psi_{12}$ represented respectively the main effects of dose and squared-dose and their respective interactions with all of the aforementioned predictors in the blip function. The treatment function underlying the blip function was specified including the main effects of all predictors $x$.

To address data privacy, we relied on the three data pooling strategies and on the distributed regression, each combined with G-dWOLS using the inverse probability of treatment weight. The first data pooling strategy used a pool size of 3 across the different sites. The second data pooling strategy used a pool size of three for all sites. The third data pooling strategy adjusted the pool size to the site in order to reduce the number of patients excluded compared to the second data pooling strategy. However, the number of pools was reduced, resulting in a loss of information. The data pooling strategies are described in Table 2. The results of the different methods were compared to the results obtained with the gold-standard method using the original individual-level data. The estimands of interest were the blip function parameters $\psi_1 = (\psi_{01}, \psi_{11})$ and $\psi_2 = (\psi_{02}, \psi_{12})$.

**Table 2.** Warfarin application: Data pooling description.

|  | Sites | | | | | | | | | Total |
|---|---|---|---|---|---|---|---|---|---|---|
|  | 2 | 4 | 5 | 8 | 11 | 12 | 14 | 15 | 21 |  |
| *Sample size* | 32 | 193 | 325 | 198 | 93 | 160 | 273 | 179 | 274 | 1727 |
| **First data pooling strategy** | | | | | | | | | | |
| *Number of pools* | | | | | | | | | | 575 |
| *Number patients excluded* | | | | | | | | | | 2 |
| **Second data pooling strategy** | | | | | | | | | | |
| *Pool size* | 3 | 3 | 3 | 3 | 3 | 3 | 3 | 3 | 3 | - |
| *Number of pools* | 10 | 64 | 108 | 66 | 31 | 53 | 91 | 59 | 91 | 573 |
| *Number patients excluded* | 2 | 1 | 1 | 0 | 0 | 1 | 0 | 2 | 1 | 8 |
| **Third data pooling strategy** | | | | | | | | | | |
| *Pool size* | 4 | 3 | 5 | 3 | 3 | 4 | 3 | 3 | 3 | - |
| *Number of pools* | 8 | 64 | 65 | 66 | 31 | 40 | 91 | 59 | 91 | 515 |
| *Number patients excluded* | 0 | 1 | 0 | 0 | 0 | 0 | 0 | 2 | 1 | 4 |

## *C. Results*

The gold-standard method using the original individual-level data being the reference, Figure 3 shows the differences between the estimates provided by this method and those provided by the data pooling strategies, and the distributed regression approach.

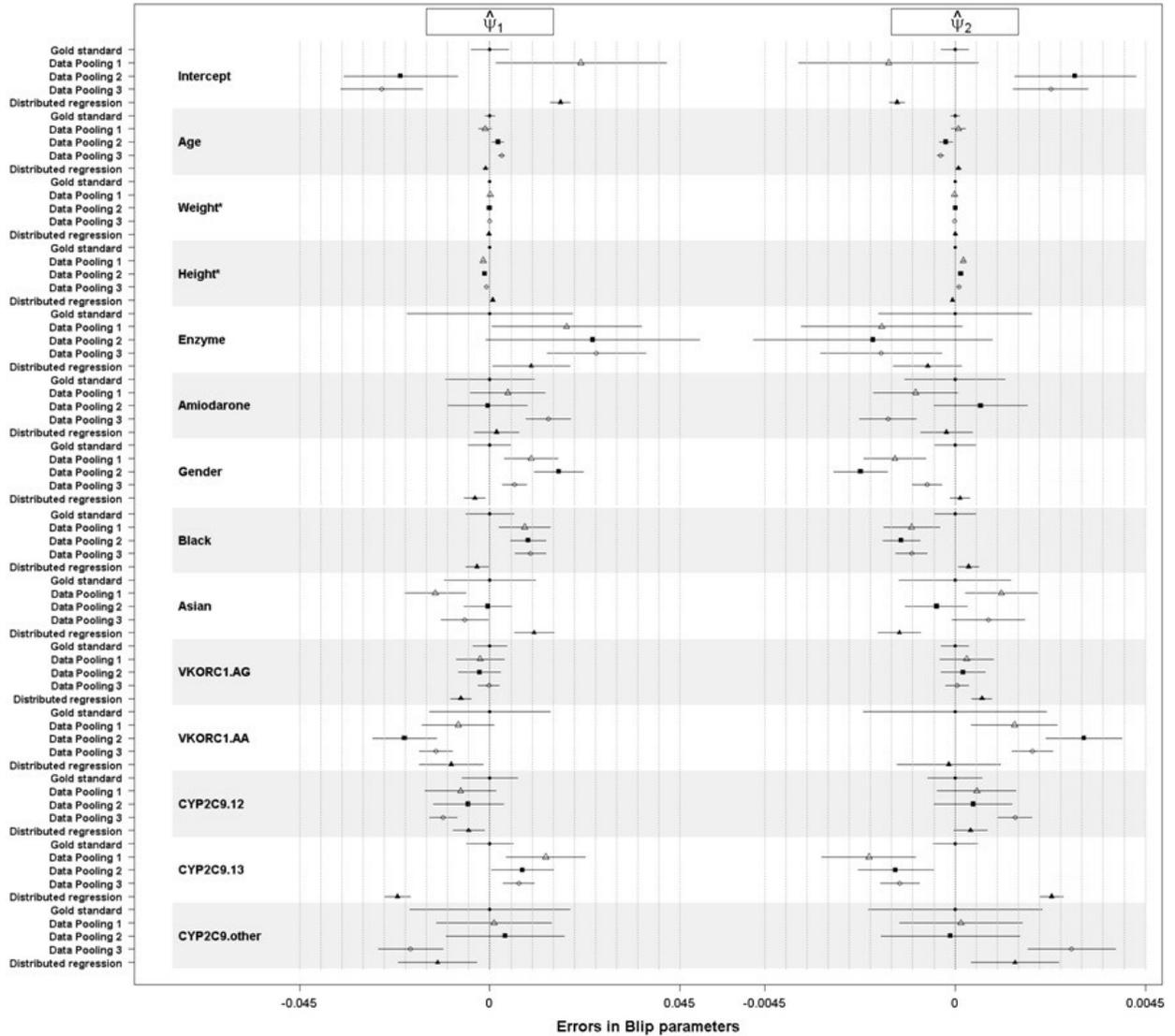

**Figure 3.** Warfarin application. Differences between the estimates provided by the gold-standard method and those provided by the data pooling strategies (Data Pooling 1, Data Pooling 2 and Data Pooling 3), and the distributed regression approach.
\* Weight and height were centered by the site mean.

Globally, we can see that distributed regression performs slightly better than the data pooling methods, in terms of errors and variability in both $\psi_1$ and $\psi_2$. The third data pooling strategy seems to perform a bit less well in terms of error than the other data pooling strategies (Figure 3), although with less variability. This may be due to the fact that less information is available, as there were 515 pools in the third strategy compared to 573 pools in the second strategy and 575 pools in the first pooling strategy, even though there were marginally fewer patients excluded in the third strategy compared to the second, 4 vs 8 patients (see Table 2).

The errors and the variability are substantially higher for $\psi_1$ than for $\psi_2$. This is likely due to the higher order interaction terms for which the scaling differs. Similar patterns have been observed in previous studies [8,20,40].

In terms of interpretation of the results, the final goal is to see if using data pooling methods or distributed regression lead to similar results compared to the gold-standard method, that is, similar estimates of the individual optimal treatment. The blip function having the following expression $\gamma(x, a; \psi) = a(\psi_{01} + \psi_{11}x + \psi_{02}a + \psi_{12}ax)$ (second part of Equation (4)), the optimal treatment, which maximizes the blip function, can be estimated as:

$$a_{opt} = -\frac{1}{2}\frac{(\psi_{01}+\psi_{11}x)}{(\psi_{02}+\psi_{12}x)},$$

with a corresponding estimate, $\hat{a}_{opt}$, obtained by substitution of estimates of $\psi$ for the parameters themselves. The rule would be then to recommend treatment $\hat{a}_{opt}$ if it is included within the range of possible values for $a$. Otherwise the optimal treatment will be that which of the maximum or minimum possible value for $a$ maximizes the estimated blip function. The optimal treatment has been estimated for each one of the five methods. The mean errors of the estimates (differences between the estimates provided by the gold-standard method and those provided by the proposed approaches) were -7.22 mg/week, 4.84 mg/week, 19.5 mg/week and 0.11 mg/week, respectively, for the first data pooling strategy, the second data pooling strategy, the third data pooling strategy and the distributed regression approach, and their absolute mean errors were respectively of 21.05 mg/week, 38.69 mg/week, 28.53 mg/week and 17.26 mg/week. Figure 4 represents the lowess (controlled by a smoother span) of the optimal treatment estimated by the five methods over the individual patients sorted by observed dose value. Clearly, the data pooling strategies provide different individual dose recommendations, probably due to the difference of information they provide (Table 2). The first data pooling strategy tends to under-estimate dose relative to the gold-standard optimal treatment, especially for higher doses. On the contrary, the second data pooling strategy tends to over-estimate the gold-standard optimal dose, especially for lower doses, with more variability. The third data pooling strategy greatly over-estimates the gold-standard optimal treatment with relatively lower variability, hence this approach showing the highest mean error but not the highest absolute mean error. The distributed regression approach globally under-estimates the gold-standard optimal treatment for the lower doses and over-estimates it beyond, with a lower variability (Figure 4). When we use a higher smoother span, we can see that distributed regression provides higher under-estimates for the lower gold-standard optimal treatment and higher over-estimates for the higher gold-standard optimal treatment. Although all these methods seem to provide different results, their confidence intervals show that these differences are not statistically significant (results not shown). A simulation study with a set-up similar to this application has also been performed (see Section 5 of the Supplementary Material). The results (Figure S3) show that data pooling 1, data pooling 2 and the distributed regression perform equivalently whereas, as noted in our application, data pooling 3 provides substantially more biased estimates.

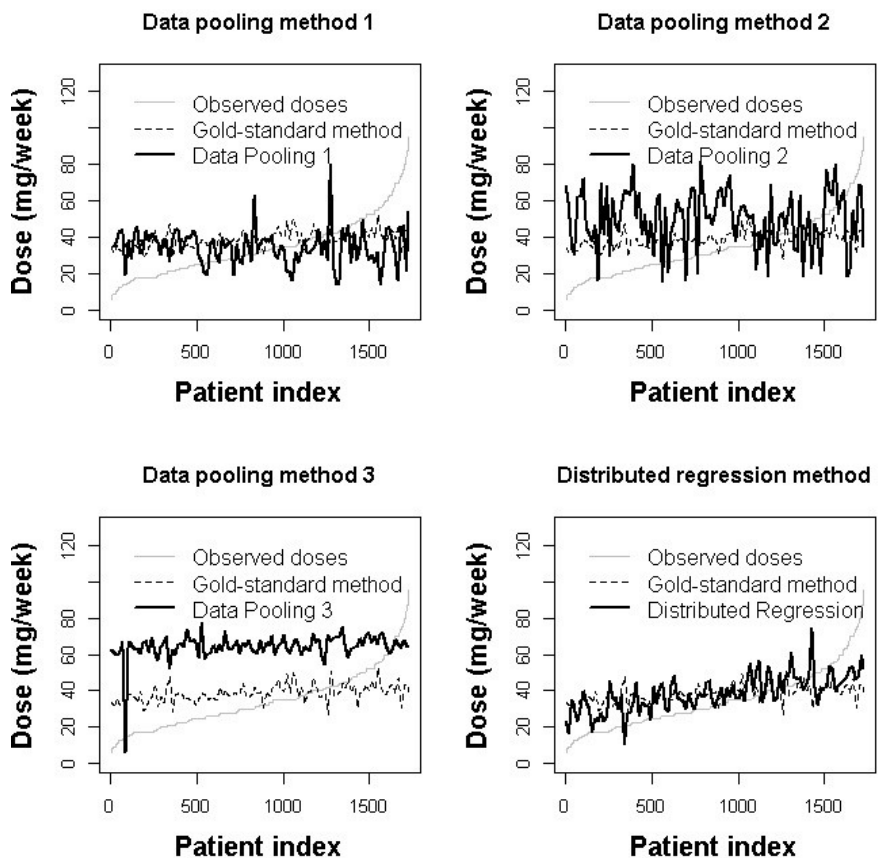

**Figure 4.** Warfarin application. Comparison of the optimal treatment estimated by the gold standard approach using individual data with no privacy preservation (solid black line) and four privacy preserving methods (the three data pooling strategies and the distributed regression approach) over the individual patients sorted by observed dose value.

## 5. Discussion

In order to provide patients with the best possible individual care while ensuring data privacy, we adapted two approaches, data pooling and distributed regression, to address concerns surrounding data sharing when attempting to use data from multiple centres to estimate optimal individualized treatment rules using G-dWOLS.

Simulations demonstrate that both methods perform well when the treatment-free model and the treatment model underlying the blip function are correct. When only the treatment-free function is correctly specified, both methods provide satisfactory results. When only the treatment function is correctly specified, the distributed regression approach performs well. Nevertheless, in the context of continuous treatment, high degrees of confounding may lead to large weights and unstable estimation and likely violate the positivity assumption. However, using the pooling data method, double robustness is not assured. Thus, to ensure doubly robustness of G-dWOLS is maintained, the distributed regression approach is superior to pooling.

In our Warfarin application aiming to estimate an optimal individualized dose strategy, we pointed out the magnitude of the errors made when the optimal individual treatment is estimated using the data pooling and the distributed regression approaches instead of the gold-standard method that uses the original individual-level data. Globally, we found that the distributed regression approach provides better results than the data pooling approaches that were used. The data pooling strategies differed from the dosing relative to the gold-standard by at least 21 mg/week on average, with some variability. The distributed regression differed for the more extreme dose recommendations though to a lesser extent. The greatest differences should be considered with interest since both over- and under-dosing can have significant adverse effects on patients. However, note that while our estimated optimal dose rule may provide some guidance for establishing a Warfarin dose algorithm, it is unlikely to be imparting a complete picture. Several important predictors were unavailable in the dataset, such as smoking status, vitamin K intake, alcohol consumption, and other genetic factors known to influence the effect of Warfarin [34].

The results for the three data pooling strategies were not presented in the simulation study since the goal of this simulation study was to see if the data pooling approach, in some form, could be adapted to address concerns of data privacy. It is reasonable to suppose that we would have obtained *broadly* similar conclusions for the two data pooling strategies that were not included in the simulation, as long as the quantity of information is equivalent. In real data, as emphasized by Saha-Chaudhuri and Weinberg [11], several pool sizes or several data pooling strategies should be used to confirm the results if this approach is pursued. However, as noted above, we would advocate the use of distributed regression over data pooling as a preferred approach.

Our encouraging results of both simulated and real data analyses, especially concerning the distributed regression approach, illustrate potential advantages in preserving data privacy when using multisite data to estimate individualized treatment rules. Indeed, in the context where the individual-level data cannot be shared, centres would need only to communicate about the covariates that should be included in each models (treatment model, treatment-free model and blip function), in order to estimate the weight matrix, needed for the main analysis that relies on weighted linear regression and to define the design matrix. Then, the centres will be able to provide the two matrix products, built from this weight matrix, this design matrix, and the outcome vector, which will allow estimating the blip parameters of interest. Concerning the data pooling method, despite less promising results, the centres would only be required to aggregate the covariates of interest according to a given pool size. The choice of the group size depends on the level of the confidentiality required and on the tolerance for loss of statistical power: when smaller group size is used, the level of the confidentiality may not be optimal but the statistical power may be higher. The choice of $g$ should balance the two issues of privacy restriction and realization of adequate statistical power. There are no other restrictions.

The methods proposed to preserve data privacy have some drawbacks. First, the data pooling strategy requires a balance between data usability and confidentiality [11,17]. Indeed, for small datasets, it is sometime difficult to create pool of decent size while not reducing the statistical power. Second, the different models (treatment model, treatment-free model and blip function) should be decided *a priori* so that the different centres can provide the corresponding aggregated covariates values when using the pooling data strategy and the corresponding weight matrix and design matrix when using the distributed regression. For any changes, centres will have to update elements that they have to provide. Analyses under the scenario where some centres do not collect data on specific covariates may be challenging. For example, multiple imputation (which has seen data-privacy adaptions [41]) would require considerable coordination between centres if a variable is entirely unavailable at one site, and the imputation model must be developed at another. And third, even if this is statistically feasible, it may not be practical to use these approaches in a multi-stage setting, as it would require going back and forth to each centre for the different steps that need to be performed at each iteration (creation of the matrix data, estimating the optimal rule, computation of the pseudo-outcome, etc...).

One might consider using traditional meta-analysis methods to combine the results obtained from each centre without needing to share individual-level data across centres. However, it has been shown that one of the limitations of traditional meta-analyses is that they are not designed for detection of subgroup effects under heterogeneity [28-30], and this could lead to very unstable estimates, since it would require combining noisy estimates. Some methods have been developed to overcome this issue, such as meta-CART [31,32,42] which integrates classification and regression trees (CART) into meta-analysis. This method can deal with many predictors and represents interactions in a parsimonious tree structure. An interesting perspective for the future would be to see if this method could be adapted to our situation, where heterogeneity in treatment response is particularly of interest

The proposed approaches to address concerns surrounding data sharing when attempting to use data from multiple centres may be of interest for other studies in precision medicine. We hope this work will stimulate further developments aimed at adapting these approaches or other privacy preserving methods to other estimation approaches for DTRs [1-10] or more generalized settings such as with censored outcomes.

**Acknowledgments**

We are very thankful to Dr Guanhua Chen (Department of Biostatistics and Medical Informatics, University of Wisconsin School of Medicine and Public Health) for kindly granting us permission to use his code used as part of our analysis of the International Warfarin Pharmacogenetics Consortium data.

# Supplementary Material: Preserving data privacy when using multi-site data to estimate individualized treatment rules

Coraline Danieli[a,*], Erica E. M. Moodie[a]

[a]McGill University, Department of Epidemiology, Biostatistics and Occupational Health

* coraline.danieli@rimuhc.ca

# 1. Summary of the simulated scenarios

Table S1 presents the 40 simulated scenarios, for which, we generated 1000 independent data sets of size $N = 60000$, divided in three centres. Using the data pooling strategy, original data were then aggregated into $N/g = 2000/600/100$ pools of size $g = 30/100/600$ patients respectively

**Table S1.** Supplementary Material: Scenarios used in the simulation study

| Scenario | Subject-specific covariates distribution | Treatment variable distribution | Level of confounding | Pool size when aggregating the data |
|---|---|---|---|---|
| a | Identical | Binary | Very low | 30 |
| b | Identical | Binary | Low | 30 |
| c | Identical | Binary | Moderate | 30 |
| d | Identical | Binary | High | 30 |
| e | Identical | Binary | Very high | 30 |
| f | Identical | Continuous | Very low | 30 |
| g | Identical | Continuous | Low | 30 |
| h | Identical | Continuous | Moderate | 30 |
| i | Identical | Continuous | High | 30 |
| j | Identical | Continuous | Very high | 30 |
| k | Different | Binary | Very low | 30 |
| l | Different | Binary | Low | 30 |
| m | Different | Binary | Moderate | 30 |
| n | Different | Binary | High | 30 |
| o | Different | Binary | Very high | 30 |
| p | Different | Continuous | Very low | 30 |
| q | Different | Continuous | Low | 30 |
| r | Different | Continuous | Moderate | 30 |
| s | Different | Continuous | High | 30 |
| t | Different | Continuous | Very high | 30 |
| a.bis | Identical | Binary | Very low | 100 |
| b.bis | Identical | Binary | Low | 100 |
| c.bis | Identical | Binary | Moderate | 100 |
| d.bis | Identical | Binary | High | 100 |
| e.bis | Identical | Binary | Very high | 100 |
| f.bis | Identical | Continuous | Very low | 100 |
| g.bis | Identical | Continuous | Low | 100 |
| h.bis | Identical | Continuous | Moderate | 100 |
| i.bis | Identical | Continuous | High | 100 |
| j.bis | Identical | Continuous | Very high | 100 |
| a.ter | Identical | Binary | Very low | 600 |
| b.ter | Identical | Binary | Low | 600 |
| c.ter | Identical | Binary | Moderate | 600 |
| d.ter | Identical | Binary | High | 600 |
| e.ter | Identical | Binary | Very high | 600 |
| f.ter | Identical | Continuous | Very low | 600 |
| g.ter | Identical | Continuous | Low | 600 |
| h.ter | Identical | Continuous | Moderate | 600 |
| i.ter | Identical | Continuous | High | 600 |
| j.ter | Identical | Continuous | Very high | 600 |

## 2. Customized program in R software to perform the simulations

The implementation was realized with R software (version 3.1.1). Below, we present the code used to generate the data and then analyze the data.

```
# -----> Data generation with continuous treatment

# correlation_level : Degree of confounding, captured by the strength of
relationship between A and X

# N : Sample size

Original_Data_ttt_cont_f <- function( correlation_level , N )
{
  if( correlation_level == "very_low" )
  {
    sd_A1 <- 12
  }
  if( correlation_level == "low" )
  {
    sd_A1 <- 4
  }
  if( correlation_level == "moderate" )
  {
```

```r
    sd_A1 <- 1.9
  }
  if( correlation_level == "high" )
  {
    sd_A1 <- 1.2
  }
  if( correlation_level == "very_high" )
  {
    sd_A1 <- 0.5
  }

  X1      <- rnorm( N , 10 , 1 )
  A1      <- rnorm( N , X1 , sd_A1 )
  epsilon <- rnorm( N )
  log_X1  <- log(X1)
  sin_X1  <- sin(X1)
  tf      <- log_X1 + sin_X1 + X1
  Y       <- log_X1 + sin_X1 + X1 + A1 * ( 1 + X1 ) + epsilon
  p1      <- rep( 0 , length(A1))

  data    <- data.frame( X1 = X1 , log_X1 = log_X1 , sin_X1 = sin_X1 , A1 = A1 , Y = Y , tf = tf , p1 = p1 )

  return( data )
}
```

```r
# -----> Data generation with binary treatment

# N : Sample size

# coeff_mult : Degree of confounding, captured by the strength of relationship between A and X

Original_Data_ttt_bin_f <- function( N , coeff_mult )

{

  expit    <- function(x) 1 / (1 + coeff_mult*exp(-(x - 10)))

  X1       <- rnorm(N , 10 , 1 )

  p1       <- expit(X1)

  A1       <- rbinom(N, 1, p1)

  epsilon <- rnorm( N )

  log_X1  <- log(X1)

  sin_X1  <- sin(X1)

  tf       <- log_X1 + sin_X1 + X1

  Y        <- log_X1 + sin_X1 + X1 + A1 * ( 1 + X1 ) + epsilon

  data     <- data.frame( X1 = X1 , log_X1 = log_X1 , sin_X1 = sin_X1 , A1 = A1 , Y = Y , tf = tf , p1 = p1)

  return( data )

}
```

# -----> Function to pool the data using the first data pooling strategy

# Original_Data : Data simulated from Original_Data_ttt_cont_f or Original_Data_ttt_bin_f function

# n : Number of pools

```r
Pooled_Data_f <- function( Original_Data , n )

{

  X1_sum     <- apply( matrix( Original_Data$X1     , ncol = n ) , 2 , sum )

  A1_sum     <- apply( matrix( Original_Data$A1     , ncol = n ) , 2 , sum )

  log_X1_sum <- apply( matrix( Original_Data$log_X1 , ncol = n ) , 2 , sum )

  sin_X1_sum <- apply( matrix( Original_Data$sin_X1 , ncol = n ) , 2 , sum )

  Y_sum      <- apply( matrix( Original_Data$Y      , ncol = n ) , 2 , sum )

  p1_sum     <- apply( matrix( Original_Data$p1     , ncol = n ) , 2 , sum )

  tf_sum     <- log_X1_sum + sin_X1_sum + X1_sum

  X1_W_sum <- c()

  for( i in 1:n )

  {

    X1_W_sum[i] <- sum( ( Original_Data$A1[((i-1)*n.pool+1):(i*n.pool)] * Original_Data$X1[((i-1)*n.pool+1):(i*n.pool)] ) / sum(Original_Data$A1[((i-1)*n.pool+1):(i*n.pool)]) )

  }
```

```r
  data <- data.frame( X1_sum = X1_sum , X1_W_sum = X1_W_sum , log_X1_sum =
log_X1_sum , sin_X1_sum = sin_X1_sum , A1_sum = A1_sum , Y_sum = Y_sum , tf_sum =
tf_sum , p1_sum = p1_sum )

  return( data )

}

# -----> Estimation of the weights when the treatment function is misspecified
for continuous treatment

# A1_t : Treatment in the treatment function

weight_norm_t0_f <- function( A1_t )

{

  ps.mod <- lm( A1_t ~ 1 )

  wt     <- 1/dnorm( A1_t , mean = ps.mod$fitted.values , sd =
summary(ps.mod)$sigma )

  return(wt)

}

# -----> Estimation of the weights when the treatment function is well specified
for continuous treatment

# A1_t : Treatment in the treatment function
```

```r
# X1_t : Covariates to adjust for in the treatment function

weight_norm_t1_f <- function( A1_t , X1_t )
{
  ps.mod <- lm( A1_t ~ 0 + X1_t )
  wt     <- 1/dnorm( A1_t , mean = ps.mod$fitted.values , sd = summary(ps.mod)$sigma )
  
  return(wt)
}

# -----> Estimation of the weights when the treatment function is misspecified for binary treatment

# A1_t : Treatment in the treatment function

weight_bin_t0_f <- function( A1_t )
{
  ps.mod <- fitted( glm( A1_t ~ 1 , family = "binomial" ) )
  wt     <- abs( A1_t - ps.mod )
  
  return(wt)
}
```

# -----> Estimation of the weights when the treatment function is well specified for binary treatment

# A1_t : Treatment in the treatment function

# X1_t : Covariates to adjust for in the treatment function

```r
weight_bin_t1_f <- function( A1_t , X1_t )
{
  ps.mod <- fitted( glm( A1_t ~ X1_t , family = "binomial" ) )
  wt     <- abs( A1_t - ps.mod )

  return(wt)
}
```

# -----> Analysis

# wt : weights estimated from weight_norm_t0_f or weight_norm_t1_f or weight_bin_t0_f or weight_bin_t1_f

# Y : Outcome

# A1_t : Treatment in the blip function

# tf : treatment-free function

# t : Covariates to adjust for in the blip function

```r
Analysis_f <- function( wt , Y , A1_t , tf , t )
{
```

```r
  est            <- lm( Y  ~ tf + A1_t + A1_t:t , weights = wt )$coef[3:4]

  b.inf          <- confint( lm( Y  ~ tf + A1_t + A1_t:t , weights = wt ) )[3:4,1]
  b.sup          <- confint( lm( Y  ~ tf + A1_t + A1_t:t , weights = wt ) )[3:4,2]

  return( list( est = est , b.inf = b.inf , b.sup = b.sup ) )

}
```

## 3. Results from simulated data where subject-specific covariate distributions do not differ across centres

Table S2 and Table S3 presents the performance of the results for scenarios a-j, in terms of relative bias and empirical standard deviation. The results were compared with results obtained from the "gold standard" analysis using the original individual-level data.

**Table S2.** Supplementary Material: Performance of the methods over 1000 samples for each scenario when the treatment distribution is BINARY and the subject-specific covariates distribution is IDENTICAL.

| Method | Degree of confounding | True $\psi_0$ | Mean $\psi_0$ | Standard deviation $\psi_0$ | Relative bias $\psi_0$ (%) |
|---|---|---|---|---|---|
| \multicolumn{6}{c}{Scenario 1 : Both models are correct} ||||||
| Gold Standard | a. Very low | 1.000 | 0.992 | 0.196 | -0.815 |
| Gold Standard | b. Low | 1.000 | 1.000 | 0.165 | -0.032 |
| Gold Standard | c. Moderate | 1.000 | 1.000 | 0.143 | 0.041 |
| Gold Standard | d. High | 1.000 | 1.005 | 0.123 | 0.472 |
| Gold Standard | e. Very high | 1.000 | 1.004 | 0.110 | 0.419 |
| Data Pooling | a. Very low | 1.000 | 0.990 | 1.219 | -0.966 |
| Data Pooling | b. Low | 1.000 | 1.016 | 0.990 | 1.571 |
| Data Pooling | c. Moderate | 1.000 | 1.014 | 0.865 | 1.355 |
| Data Pooling | d. High | 1.000 | 1.004 | 0.742 | 0.395 |
| Data Pooling | e. Very high | 1.000 | 0.976 | 0.680 | -2.387 |
| Distributed regression | a. Very low | 1.000 | 0.992 | 0.196 | -0.821 |
| Distributed regression | b. Low | 1.000 | 1.000 | 0.165 | -0.026 |
| Distributed regression | c. Moderate | 1.000 | 1.001 | 0.143 | 0.062 |
| Distributed regression | d. High | 1.000 | 1.005 | 0.123 | 0.488 |
| Distributed regression | e. Very high | 1.000 | 1.004 | 0.110 | 0.415 |
| \multicolumn{6}{c}{Scenario 2 : Treatment model is correct} ||||||
| Gold Standard | a. Very low | 1.000 | 0.993 | 0.202 | -0.747 |
| Gold Standard | b. Low | 1.000 | 1.000 | 0.166 | 0.034 |
| Gold Standard | c. Moderate | 1.000 | 0.999 | 0.146 | -0.055 |
| Gold Standard | d. High | 1.000 | 1.002 | 0.127 | 0.209 |
| Gold Standard | e. Very high | 1.000 | 1.004 | 0.118 | 0.436 |
| Data Pooling | a. Very low | 1.000 | -3.365 | 1.215 | -436.519 |
| Data Pooling | b. Low | 1.000 | -3.035 | 0.980 | -403.541 |
| Data Pooling | c. Moderate | 1.000 | -2.770 | 0.837 | -377.008 |
| Data Pooling | d. High | 1.000 | -2.219 | 0.683 | -321.864 |
| Data Pooling | e. Very high | 1.000 | -1.807 | 0.666 | -280.706 |
| Distributed regression | a. Very low | 1.000 | 0.992 | 0.202 | -0.809 |
| Distributed regression | b. Low | 1.000 | 1.000 | 0.166 | 0.016 |
| Distributed regression | c. Moderate | 1.000 | 0.999 | 0.146 | -0.100 |
| Distributed regression | d. High | 1.000 | 1.002 | 0.127 | 0.202 |
| Distributed regression | e. Very high | 1.000 | 1.004 | 0.118 | 0.436 |
| \multicolumn{6}{c}{Scenario 3 : Treatment-free model is correct} ||||||
| Gold Standard | a. Very low | 1.000 | 0.990 | 0.198 | -1.027 |
| Gold Standard | b. Low | 1.000 | 0.999 | 0.160 | -0.132 |
| Gold Standard | c. Moderate | 1.000 | 1.000 | 0.135 | 0.009 |
| Gold Standard | d. High | 1.000 | 1.005 | 0.113 | 0.471 |
| Gold Standard | e. Very high | 1.000 | 1.005 | 0.098 | 0.455 |
| Data Pooling | a. Very low | 1.000 | 0.997 | 1.222 | -0.301 |
| Data Pooling | b. Low | 1.000 | 1.011 | 0.975 | 1.077 |
| Data Pooling | c. Moderate | 1.000 | 0.996 | 0.865 | -0.364 |
| Data Pooling | d. High | 1.000 | 1.008 | 0.739 | 0.827 |
| Data Pooling | e. Very high | 1.000 | 0.975 | 0.677 | -2.457 |
| Distributed regression | a. Very low | 1.000 | 0.990 | 0.198 | -1.017 |
| Distributed regression | b. Low | 1.000 | 0.999 | 0.160 | -0.138 |
| Distributed regression | c. Moderate | 1.000 | 1.000 | 0.136 | 0.004 |
| Distributed regression | d. High | 1.000 | 1.005 | 0.113 | 0.460 |
| Distributed regression | e. Very high | 1.000 | 1.005 | 0.098 | 0.451 |
| \multicolumn{6}{c}{Scenario 4 : Neither model is correct} ||||||
| Gold Standard | a. Very low | 1.000 | -3.508 | 0.201 | -450.793 |
| Gold Standard | b. Low | 1.000 | -3.186 | 0.156 | -418.637 |
| Gold Standard | c. Moderate | 1.000 | -2.917 | 0.136 | -391.714 |
| Gold Standard | d. High | 1.000 | -2.329 | 0.115 | -332.920 |
| Gold Standard | e. Very high | 1.000 | -1.855 | 0.106 | -285.528 |
| Data Pooling | a. Very low | 1.000 | -3.500 | 1.219 | -449.980 |
| Data Pooling | b. Low | 1.000 | -3.171 | 0.969 | -417.119 |
| Data Pooling | c. Moderate | 1.000 | -2.907 | 0.848 | -390.653 |
| Data Pooling | d. High | 1.000 | -2.307 | 0.693 | -330.747 |
| Data Pooling | e. Very high | 1.000 | -1.874 | 0.662 | -287.427 |
| Distributed regression | a. Very low | 1.000 | -3.508 | 0.201 | -450.790 |
| Distributed regression | b. Low | 1.000 | -3.186 | 0.157 | -418.636 |
| Distributed regression | c. Moderate | 1.000 | -2.917 | 0.136 | -391.699 |
| Distributed regression | d. High | 1.000 | -2.329 | 0.115 | -332.916 |
| Distributed regression | e. Very high | 1.000 | -1.855 | 0.106 | -285.538 |

**Table S3.** Supplementary Material: Performance of the methods over 1000 samples for each scenario when the treatment distribution is CONTINUOUS and the subject-specific covariates distribution is IDENTICAL.

| Method | Degree of confounding | True $\psi_0$ | Mean $\psi_0$ | Standard deviation $\psi_0$ | Relative bias $\psi_0$ (%) |
|---|---|---|---|---|---|
| | | Scenario 1 : Both models are correct | | | |
| Gold Standard | f. Very low | 1.000 | 1.001 | 0.041 | 0.070 |
| Gold Standard | g. Low | 1.000 | 1.001 | 0.108 | 0.108 |
| Gold Standard | h. Moderate | 1.000 | 0.991 | 0.167 | -0.950 |
| Gold Standard | i. High | 1.000 | 1.002 | 0.207 | 0.177 |
| Gold Standard | j. Very high | 1.000 | 1.011 | 0.345 | 1.125 |
| Data Pooling | f. Very low | 1.000 | 1.004 | 0.086 | 0.383 |
| Data Pooling | g. Low | 1.000 | 0.987 | 0.205 | -1.295 |
| Data Pooling | h. Moderate | 1.000 | 1.012 | 0.277 | 1.249 |
| Data Pooling | i. High | 1.000 | 0.998 | 0.298 | -0.160 |
| Data Pooling | j. Very high | 1.000 | 1.008 | 0.467 | 0.777 |
| Distributed regression | f. Very low | 1.000 | 1.001 | 0.041 | 0.060 |
| Distributed regression | g. Low | 1.000 | 1.001 | 0.108 | 0.109 |
| Distributed regression | h. Moderate | 1.000 | 0.990 | 0.166 | -0.987 |
| Distributed regression | i. High | 1.000 | 1.001 | 0.207 | 0.140 |
| Distributed regression | j. Very high | 1.000 | 1.010 | 0.343 | 1.036 |
| | | Scenario 2 : Treatment model is correct | | | |
| Gold Standard | f. Very low | 1.000 | 0.998 | 0.044 | -0.197 |
| Gold Standard | g. Low | 1.000 | 0.971 | 0.127 | -2.860 |
| Gold Standard | h. Moderate | 1.000 | 0.859 | 0.235 | -14.065 |
| Gold Standard | i. High | 1.000 | 0.724 | 0.368 | -27.624 |
| Gold Standard | j. Very high | 1.000 | 0.124 | 0.602 | -87.640 |
| Data Pooling | f. Very low | 1.000 | 0.986 | 0.094 | -1.437 |
| Data Pooling | g. Low | 1.000 | 0.831 | 0.292 | -16.901 |
| Data Pooling | h. Moderate | 1.000 | 0.496 | 0.507 | -50.357 |
| Data Pooling | i. High | 1.000 | 0.094 | 0.622 | -90.555 |
| Data Pooling | j. Very high | 1.000 | -0.406 | 0.805 | -140.571 |
| Distributed regression | f. Very low | 1.000 | 0.998 | 0.045 | -0.196 |
| Distributed regression | g. Low | 1.000 | 0.971 | 0.128 | -2.860 |
| Distributed regression | h. Moderate | 1.000 | 0.859 | 0.235 | -14.104 |
| Distributed regression | i. High | 1.000 | 0.724 | 0.367 | -27.627 |
| Distributed regression | j. Very high | 1.000 | 0.124 | 0.599 | -87.596 |
| | | Scenario 3 : Treatment-free model is correct | | | |
| Gold Standard | f. Very low | 1.000 | 1.000 | 0.042 | 0.039 |
| Gold Standard | g. Low | 1.000 | 1.003 | 0.117 | 0.251 |
| Gold Standard | h. Moderate | 1.000 | 0.998 | 0.194 | -0.209 |
| Gold Standard | i. High | 1.000 | 1.016 | 0.230 | 1.579 |
| Gold Standard | j. Very high | 1.000 | 1.002 | 0.329 | 0.232 |
| Data Pooling | f. Very low | 1.000 | 1.002 | 0.087 | 0.249 |
| Data Pooling | g. Low | 1.000 | 0.983 | 0.246 | -1.703 |
| Data Pooling | h. Moderate | 1.000 | 1.009 | 0.295 | 0.851 |
| Data Pooling | i. High | 1.000 | 0.995 | 0.316 | -0.490 |
| Data Pooling | j. Very high | 1.000 | 0.989 | 0.550 | -1.050 |
| Distributed regression | f. Very low | 1.000 | 1.000 | 0.043 | 0.031 |
| Distributed regression | g. Low | 1.000 | 1.002 | 0.118 | 0.250 |
| Distributed regression | h. Moderate | 1.000 | 0.997 | 0.194 | -0.281 |
| Distributed regression | i. High | 1.000 | 1.016 | 0.229 | 1.598 |
| Distributed regression | j. Very high | 1.000 | 1.002 | 0.329 | 0.234 |
| | | Scenario 4 : Neither model is correct | | | |
| Gold Standard | f. Very low | 1.000 | 0.980 | 0.044 | -2.022 |
| Gold Standard | g. Low | 1.000 | 0.850 | 0.124 | -14.980 |
| Gold Standard | h. Moderate | 1.000 | 0.612 | 0.219 | -38.840 |
| Gold Standard | i. High | 1.000 | 0.490 | 0.305 | -51.029 |
| Gold Standard | j. Very high | 1.000 | 0.422 | 0.372 | -57.813 |
| Data Pooling | f. Very low | 1.000 | 0.983 | 0.094 | -1.734 |
| Data Pooling | g. Low | 1.000 | 0.812 | 0.313 | -18.825 |
| Data Pooling | h. Moderate | 1.000 | 0.468 | 0.529 | -53.190 |
| Data Pooling | i. High | 1.000 | 0.087 | 0.687 | -91.323 |
| Data Pooling | j. Very high | 1.000 | -0.376 | 0.866 | -137.580 |
| Distributed regression | f. Very low | 1.000 | 0.980 | 0.044 | -2.028 |
| Distributed regression | g. Low | 1.000 | 0.850 | 0.124 | -14.984 |
| Distributed regression | h. Moderate | 1.000 | 0.611 | 0.219 | -38.920 |
| Distributed regression | i. High | 1.000 | 0.490 | 0.306 | -50.986 |
| Distributed regression | j. Very high | 1.000 | 0.422 | 0.371 | -57.812 |

## 4. Results from simulated data where subject-specific covariate distributions differ across centres

The effect of varying the distribution of the subject-specific covariate over centres has been investigated. Whether treatment is binary (Table S4, Figure S1) or continuous (Table S5 and Figure S2), the results were broadly similar to those above, observed when the distribution of the subject-specific covariate is identical (results described in the main manuscript). The exceptions are that the data pooling performed rather more poorly than the individual-level data and distributed regression in the binary data setting, and, in the continuous data setting with Scenario 3 and a high degree of confounding, considerable instability in estimation with distributed regression was observed.

**Table S4.** Supplementary Material. Performance of the method over 1000 samples for each scenario when the treatment distribution is BINARY and the subject-specific covariates distribution is DIFFERENT.

| Method | Degree of confounding | True $\psi_0$ | Mean $\psi_0$ | Standard deviation $\psi_0$ | Relative bias $\psi_0$ (%) |
|---|---|---|---|---|---|
| *Scenario 1 : Both models are correct* | | | | | |
| Gold Standard | k. Very low | 1.000 | 1.001 | 0.135 | 0.133 |
| Gold Standard | l. Low | 1.000 | 1.003 | 0.123 | 0.328 |
| Gold Standard | m. Moderate | 1.000 | 1.001 | 0.114 | 0.078 |
| Gold Standard | n. High | 1.000 | 0.997 | 0.096 | -0.284 |
| Gold Standard | o. Very high | 1.000 | 0.997 | 0.087 | -0.339 |
| Data Pooling | k. Very low | 1.000 | 0.953 | 0.622 | -4.729 |
| Data Pooling | l. Low | 1.000 | 0.958 | 0.475 | -4.216 |
| Data Pooling | m. Moderate | 1.000 | 0.958 | 0.396 | -4.161 |
| Data Pooling | n. High | 1.000 | 0.984 | 0.319 | -1.603 |
| Data Pooling | o. Very high | 1.000 | 0.989 | 0.282 | -1.086 |
| Distributed regression | k. Very low | 1.000 | 1.001 | 0.135 | 0.130 |
| Distributed regression | l. Low | 1.000 | 1.003 | 0.123 | 0.325 |
| Distributed regression | m. Moderate | 1.000 | 1.001 | 0.114 | 0.088 |
| Distributed regression | n. High | 1.000 | 0.997 | 0.096 | -0.289 |
| Distributed regression | o. Very high | 1.000 | 0.996 | 0.087 | -0.356 |
| *Scenario 2 : Treatment model is correct* | | | | | |
| Gold Standard | k. Very low | 1.000 | 1.005 | 0.143 | 0.494 |
| Gold Standard | l. Low | 1.000 | 1.006 | 0.125 | 0.574 |
| Gold Standard | m. Moderate | 1.000 | 1.001 | 0.116 | 0.071 |
| Gold Standard | n. High | 1.000 | 0.996 | 0.110 | -0.409 |
| Gold Standard | o. Very high | 1.000 | 0.994 | 0.112 | -0.602 |
| Data Pooling | k. Very low | 1.000 | -7.745 | 0.742 | -874.522 |
| Data Pooling | l. Low | 1.000 | -6.831 | 0.531 | -783.078 |
| Data Pooling | m. Moderate | 1.000 | -6.173 | 0.431 | -717.331 |
| Data Pooling | n. High | 1.000 | -5.167 | 0.300 | -616.744 |
| Data Pooling | o. Very high | 1.000 | -4.826 | 0.252 | -582.556 |
| Distributed regression | k. Very low | 1.000 | 1.004 | 0.141 | 0.391 |
| Distributed regression | l. Low | 1.000 | 1.005 | 0.124 | 0.483 |
| Distributed regression | m. Moderate | 1.000 | 1.001 | 0.115 | 0.083 |
| Distributed regression | n. High | 1.000 | 0.996 | 0.108 | -0.430 |
| Distributed regression | o. Very high | 1.000 | 0.994 | 0.110 | -0.608 |
| *Scenario 3 : Treatment-free model is correct* | | | | | |
| Gold Standard | k. Very low | 1.000 | 0.999 | 0.113 | -0.124 |
| Gold Standard | l. Low | 1.000 | 1.000 | 0.098 | 0.001 |
| Gold Standard | m. Moderate | 1.000 | 0.998 | 0.090 | -0.163 |
| Gold Standard | n. High | 1.000 | 0.996 | 0.074 | -0.392 |
| Gold Standard | o. Very high | 1.000 | 0.995 | 0.066 | -0.468 |
| Data Pooling | k. Very low | 1.000 | 0.957 | 0.632 | -4.274 |
| Data Pooling | l. Low | 1.000 | 0.966 | 0.493 | -3.445 |
| Data Pooling | m. Moderate | 1.000 | 0.963 | 0.410 | -3.654 |
| Data Pooling | n. High | 1.000 | 0.983 | 0.318 | -1.740 |
| Data Pooling | o. Very high | 1.000 | 0.993 | 0.283 | -0.703 |
| Distributed regression | k. Very low | 1.000 | 0.999 | 0.110 | -0.143 |
| Distributed regression | l. Low | 1.000 | 1.000 | 0.095 | -0.023 |
| Distributed regression | m. Moderate | 1.000 | 0.998 | 0.088 | -0.174 |
| Distributed regression | n. High | 1.000 | 0.996 | 0.074 | -0.380 |
| Distributed regression | o. Very high | 1.000 | 0.996 | 0.067 | -0.439 |
| *Scenario 4 : Neither model is correct* | | | | | |
| Gold Standard | k. Very low | 1.000 | -4.789 | 0.138 | -578.862 |
| Gold Standard | l. Low | 1.000 | -4.587 | 0.112 | -558.697 |
| Gold Standard | m. Moderate | 1.000 | -4.295 | 0.099 | -529.547 |
| Gold Standard | n. High | 1.000 | -3.400 | 0.082 | -439.998 |
| Gold Standard | o. Very high | 1.000 | -2.497 | 0.079 | -349.735 |
| Data Pooling | k. Very low | 1.000 | -9.536 | 0.726 | -1053.569 |
| Data Pooling | l. Low | 1.000 | -8.409 | 0.515 | -940.875 |
| Data Pooling | m. Moderate | 1.000 | -7.619 | 0.405 | -861.908 |
| Data Pooling | n. High | 1.000 | -6.341 | 0.278 | -734.058 |
| Data Pooling | o. Very high | 1.000 | -5.536 | 0.245 | -653.560 |
| Distributed regression | k. Very low | 1.000 | -3.909 | 0.137 | -490.857 |
| Distributed regression | l. Low | 1.000 | -3.770 | 0.112 | -476.967 |
| Distributed regression | m. Moderate | 1.000 | -3.548 | 0.098 | -454.846 |
| Distributed regression | n. High | 1.000 | -2.841 | 0.081 | -384.083 |
| Distributed regression | o. Very high | 1.000 | -2.120 | 0.078 | -311.990 |

**Table S5.** Supplementary Material. Performance of the method over 1000 samples* for each scenario when the treatment distribution is CONTINUOUS and the subject-specific covariates distribution is DIFFERENT.

*Due to instability because of colinearity issue, the distributed regression method did not converge for 54, 222 and 521 samples, respectively for scenario 3-r, 3-s and 3-t and for 55, 270 and 649 samples, respectively for scenario 4-r, 4-s and 4-t.

| Method | Degree of confounding | True $\psi_0$ | Mean $\psi_0$ | Standard deviation $\psi_0$ | Relative bias $\psi_0$ (%) |
|---|---|---|---|---|---|
| Scenario 1 : Both models are correct ||||||
| Gold Standard | p. Very low | 1.000 | 1.000 | 0.024 | -0.043 |
| Gold Standard | q. Low | 1.000 | 0.998 | 0.066 | -0.179 |
| Gold Standard | r. Moderate | 1.000 | 0.999 | 0.123 | -0.053 |
| Gold Standard | s. High | 1.000 | 0.998 | 0.147 | -0.240 |
| Gold Standard | t. Very high | 1.000 | 1.012 | 0.229 | 1.226 |
| Data Pooling | p. Very low | 1.000 | 1.000 | 0.049 | -0.027 |
| Data Pooling | q. Low | 1.000 | 1.007 | 0.089 | 0.707 |
| Data Pooling | r. Moderate | 1.000 | 1.006 | 0.132 | 0.550 |
| Data Pooling | s. High | 1.000 | 1.007 | 0.171 | 0.699 |
| Data Pooling | t. Very high | 1.000 | 1.007 | 0.267 | 0.682 |
| Distributed regression | p. Very low | 1.000 | 1.000 | 0.024 | -0.046 |
| Distributed regression | q. Low | 1.000 | 0.998 | 0.066 | -0.170 |
| Distributed regression | r. Moderate | 1.000 | 0.999 | 0.123 | -0.067 |
| Distributed regression | s. High | 1.000 | 0.998 | 0.147 | -0.236 |
| Distributed regression | t. Very high | 1.000 | 1.012 | 0.228 | 1.150 |
| Scenario 2 : Treatment model is correct ||||||
| Gold Standard | p. Very low | 1.000 | 0.995 | 0.028 | -0.461 |
| Gold Standard | q. Low | 1.000 | 0.963 | 0.081 | -3.714 |
| Gold Standard | r. Moderate | 1.000 | 0.859 | 0.151 | -14.080 |
| Gold Standard | s. High | 1.000 | 0.734 | 0.199 | -26.648 |
| Gold Standard | t. Very high | 1.000 | 0.436 | 0.271 | -56.365 |
| Data Pooling | p. Very low | 1.000 | 0.843 | 0.074 | -15.748 |
| Data Pooling | q. Low | 1.000 | 0.448 | 0.128 | -55.198 |
| Data Pooling | r. Moderate | 1.000 | 0.263 | 0.150 | -73.666 |
| Data Pooling | s. High | 1.000 | 0.225 | 0.169 | -77.485 |
| Data Pooling | t. Very high | 1.000 | 0.194 | 0.276 | -80.567 |
| Distributed regression | p. Very low | 1.000 | 0.996 | 0.029 | -0.447 |
| Distributed regression | q. Low | 1.000 | 0.963 | 0.081 | -3.691 |
| Distributed regression | r. Moderate | 1.000 | 0.859 | 0.150 | -14.117 |
| Distributed regression | s. High | 1.000 | 0.733 | 0.199 | -26.666 |
| Distributed regression | t. Very high | 1.000 | 0.437 | 0.271 | -56.340 |
| Scenario 3 : Treatment-free model is correct ||||||
| Gold Standard | p. Very low | 1.000 | 1.000 | 0.023 | -0.043 |
| Gold Standard | q. Low | 1.000 | 1.001 | 0.056 | 0.103 |
| Gold Standard | r. Moderate | 1.000 | 0.996 | 0.063 | -0.400 |
| Gold Standard | s. High | 1.000 | 0.975 | 0.156 | -2.500 |
| Gold Standard | t. Very high | 1.000 | 0.930 | 0.255 | -7.000 |
| Data Pooling | p. Very low | 1.000 | 1.000 | 0.047 | 0.014 |
| Data Pooling | q. Low | 1.000 | 1.002 | 0.087 | 0.239 |
| Data Pooling | r. Moderate | 1.000 | 1.004 | 0.094 | 0.358 |
| Data Pooling | s. High | 1.000 | 0.996 | 0.125 | -0.359 |
| Data Pooling | t. Very high | 1.000 | 1.002 | 0.163 | 0.175 |
| Distributed regression | p. Very low | 1.000 | 1.000 | 0.023 | -0.032 |
| Distributed regression | q. Low | 1.000 | 1.002 | 0.060 | 0.203 |
| Distributed regression | r. Moderate | 1.000 | 0.979 | 0.559 | -2.130 |
| Distributed regression | s. High | 1.000 | 1.115 | 4.035 | 11.469 |
| Distributed regression | t. Very high | 1.000 | 0.534 | 20.421 | -46.624 |
| Scenario 4 : Neither model is correct ||||||
| Gold Standard | p. Very low | 1.000 | 0.975 | 0.028 | -2.526 |
| Gold Standard | q. Low | 1.000 | 0.911 | 0.070 | -8.859 |
| Gold Standard | r. Moderate | 1.000 | 0.996 | 0.063 | -0.400 |
| Gold Standard | s. High | 1.000 | 0.974 | 0.159 | -2.600 |
| Gold Standard | t. Very high | 1.000 | 0.928 | 0.259 | -7.200 |
| Data Pooling | p. Very low | 1.000 | 0.813 | 0.070 | -18.728 |
| Data Pooling | q. Low | 1.000 | 0.367 | 0.142 | -63.293 |
| Data Pooling | r. Moderate | 1.000 | 0.143 | 0.134 | -85.654 |
| Data Pooling | s. High | 1.000 | 0.067 | 0.140 | -93.321 |
| Data Pooling | t. Very high | 1.000 | 0.016 | 0.249 | -98.355 |
| Distributed regression | p. Very low | 1.000 | 0.977 | 0.028 | -2.330 |
| Distributed regression | q. Low | 1.000 | 0.909 | 0.067 | -9.115 |
| Distributed regression | r. Moderate | 1.000 | 0.930 | 0.861 | -7.044 |
| Distributed regression | s. High | 1.000 | 1.889 | 12.443 | 88.932 |
| Distributed regression | t. Very high | 1.000 | -2.362 | 116.592 | -336.205 |

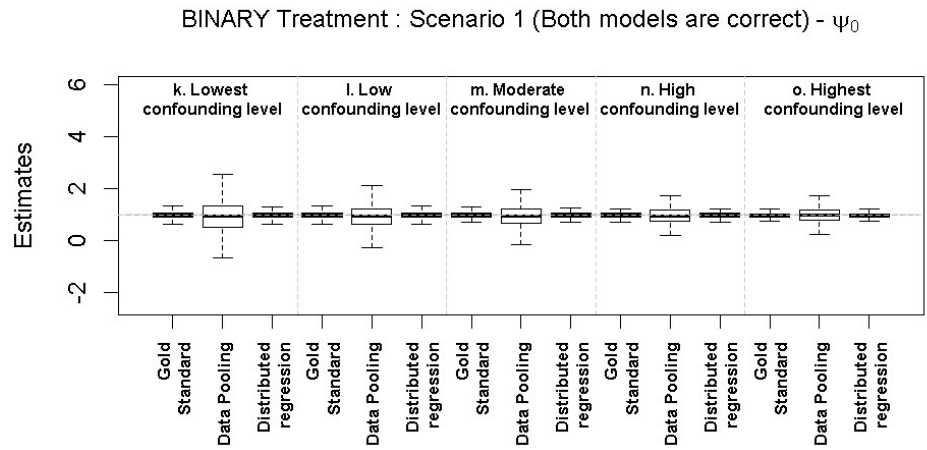
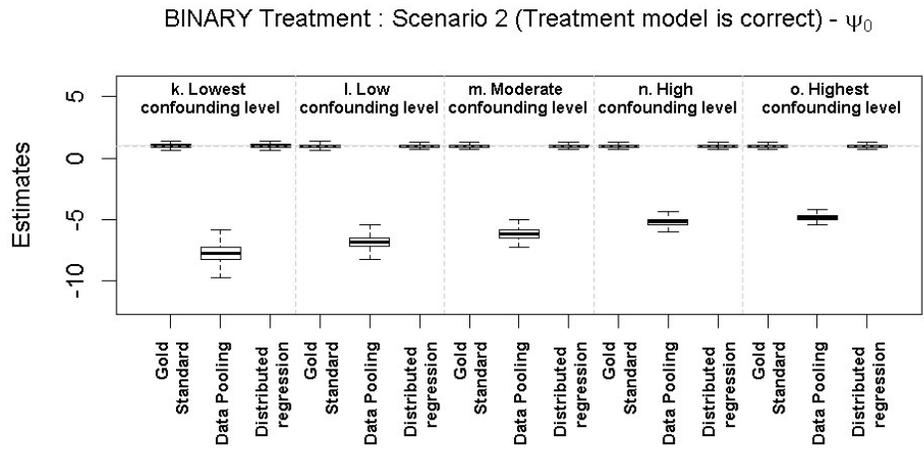
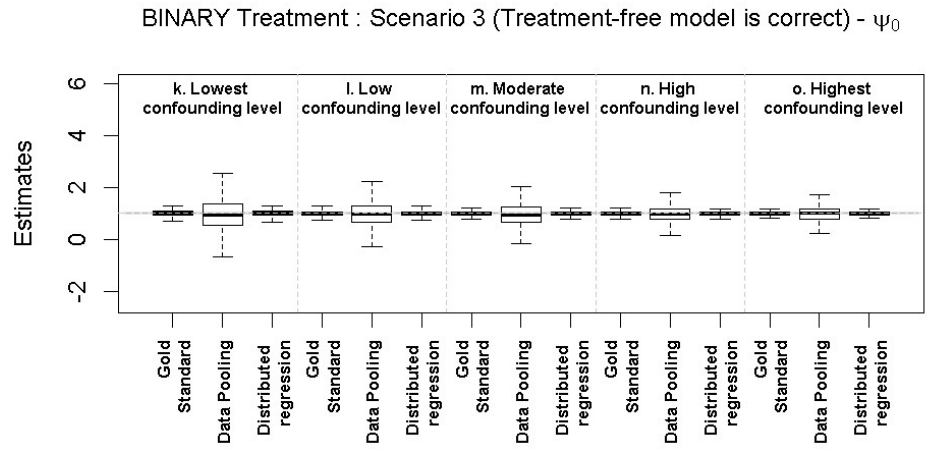
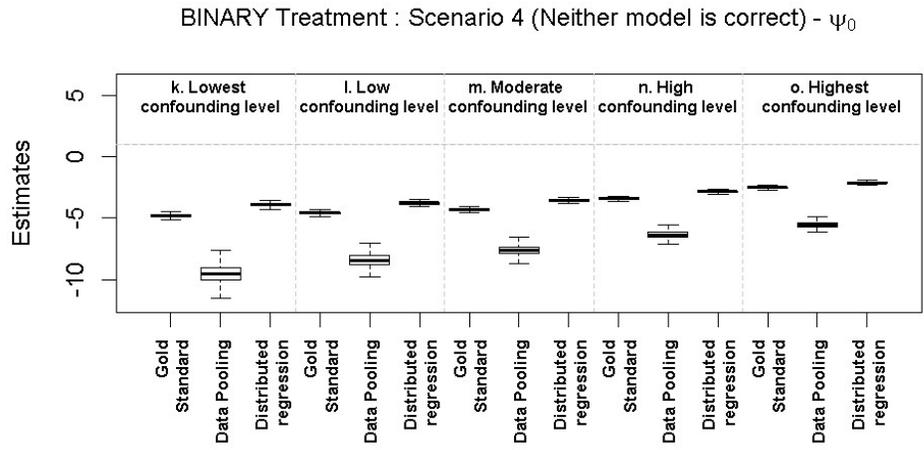

**Figure S1.** Supplementary Material. Performance of the method over 1000 samples for each scenario when the treatment distribution is BINARY and the subject-specific covariates distribution is DIFFERENT. (The pool size when aggregating the data is 30).

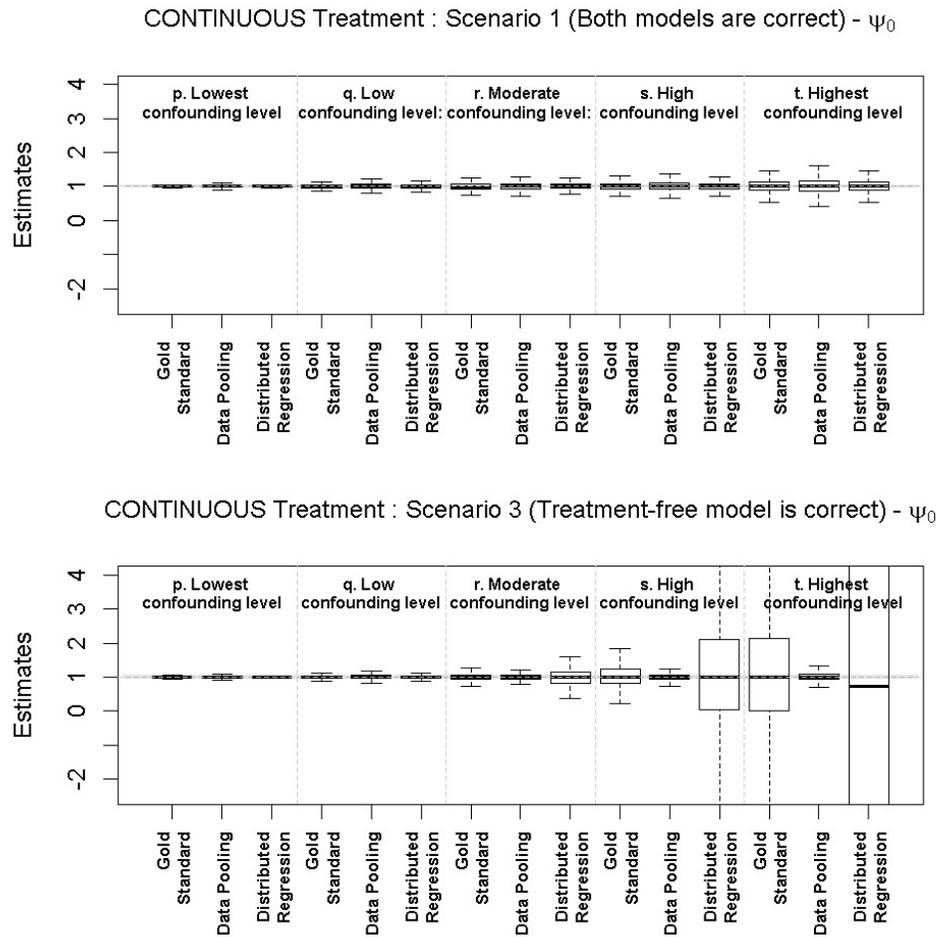
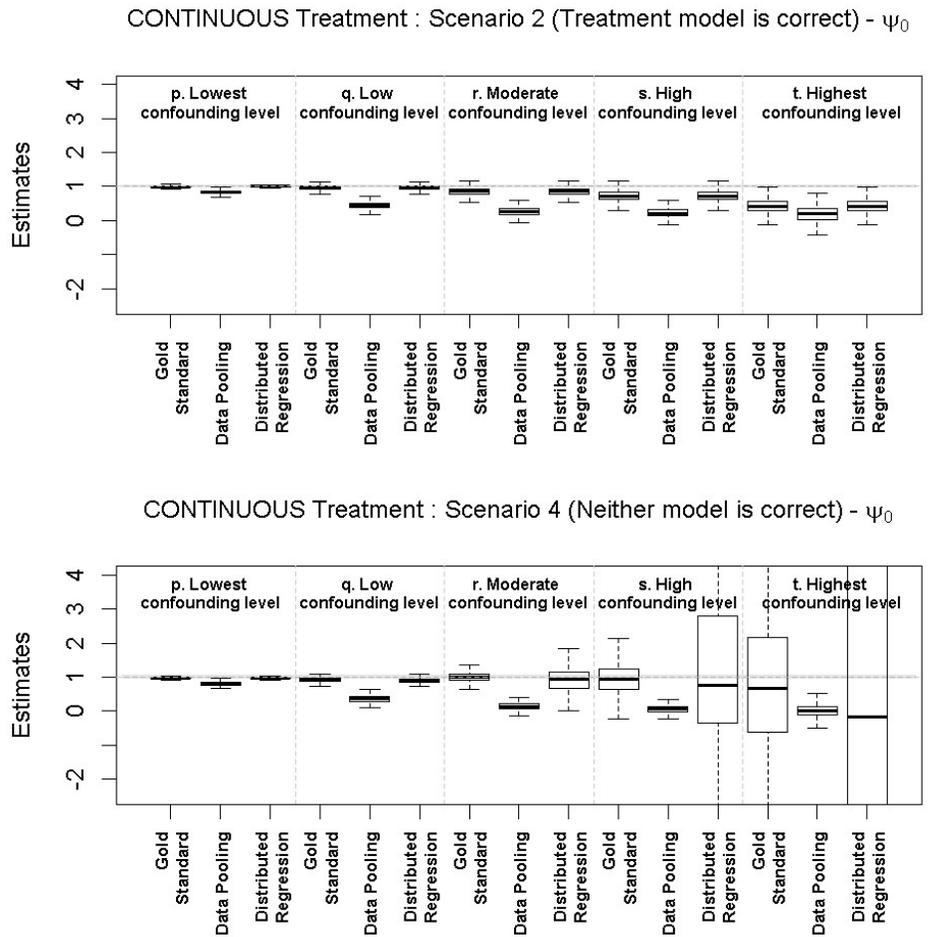

**Figure S2** Supplementary Material. Performance of the method over 1000 samples* for each scenario when the treatment distribution is CONTINOUS and the subject-specific covariates distribution is DIFFERENT. Note that some the y-axis has been restricted to better show variation in the width of the "boxes" at the cost of being unable to see the entire boxplot in the settings where confounding is at its highest levels in scenarios 3 and 4. (The pool size when aggregating the data is 30).

*Due to instability because of colinearity issue, the distributed regression method did not converge for 54, 222 and 521 samples, respectively for scenario 3-r, 3-s and 3-t and for 55, 270 and 649 samples, respectively for scenario 4-r, 4-s and 4-t.

## 5. Simulation study with a set-up similar to the application on estimating an optimal warfarin dose strategy

As in the main text, this simulation study is reported following the ADEMP (aims, data-generating mechanisms, estimands, methods, and performance measures) scheme[32].

### A. Aims

The objective is to assess and compare the performance of the proposed methods within a set-up similar to the application on estimating an optimal warfarin dose strategy.

### B. Data-generating mechanisms

We generated 300 independent data sets of size $N = 1727$, divided among 9 sites whose respective sample sizes are as in those reported in Table 2 of the main document. For each site, we generated the predictors (patient's age, the weight, the height, an indicator for taking either Rifampin, Carbamazepine or Phenytoin, an indicator for taking Amiodarone, gender, race, and genetic information on the VKORC1 and CYP2C9 genotypes; these covariates all form the vector $x^{sim}$) according to the observed distributions in the real dataset (normal distribution for continuous covariates with specific parameters (mean and standard error) and Bernoulli distribution for binary covariate with a specific parameter (probability of success)). Using the parameters $\widehat{\beta}^{treatment}$ obtained from the treatment model on the observed data (the treatment function underlying the blip function was specified including the main effects of all predictors) and the simulated predictors $x^{sim}$, the treatment $a^{sim}$ was generated using a normal distribution $N(\mu, \sigma)$ with $\mu = E(\widehat{\beta}^{treatment} x^{sim})$ and $\sigma = SD(\widehat{\beta}^{treatment} x^{sim})$. In the same way, using the parameters $(\widehat{\beta}, \widehat{\psi}_1, \widehat{\psi}_2)$ obtained from the outcome model on the observed data, the simulated predictors $x^{sim}$ and the simulated treatment $a^{sim}$, the outcome was generated using a normal distribution $N(\mu, \sigma)$ with

$$\mu = E\left(x^{sim}\widehat{\beta} + a^{sim}(\widehat{\psi}_{01} + x^{sim}\widehat{\psi}_{11} + a^{sim}\widehat{\psi}_{02} + a^{sim}x^{sim}\widehat{\psi}_{12})\right)$$

and

$$\sigma = SD\left(x^{sim}\widehat{\beta} + a^{sim}(\widehat{\psi}_{01} + x^{sim}\widehat{\psi}_{11} + a^{sim}\widehat{\psi}_{02} + a^{sim}x^{sim}\widehat{\psi}_{12})\right).$$

All simulations were performed using a customized program in R.

## C. Estimands, methods, and performance metrics

The analyses using the G-dWOLS method relied on a blip function that is quadratic in the Warfarin dose, including the main effects of both dose and squared-dose as well as their respective interactions with all of the aforementioned predictors ($x^{\psi} = x^{sim}$). Thus, in accordance with the real application, the outcome model was specified as:

$$y^{sim} = x^{sim}\beta + a^{sim}(\psi_{01} + x^{sim}\psi_{11} + a^{sim}\psi_{02} + a^{sim}x^{sim}\psi_{12}). \tag{S1}$$

The vector $\beta$ represented the effect of all predictors $x^{sim}$ that specified the treatment-free function ($x^{sim,\beta} = x^{sim}$), on the outcome $y^{sim}$. The vector $a^{sim}$ represented the vector of dose, and $\psi_{01}, \psi_{02}, \psi_{11}$ and $\psi_{12}$ represented respectively the main effects of dose and squared-dose and their respective interactions with all of the aforementioned predictors in the blip function. The treatment function underlying the blip function was specified including the main effects of all predictors $x^{sim}$.

To address data privacy, we relied on all data pooling strategy, which allowed using several pool sizes as recommended by Saha-Chaudhuri and Weinberg[11], and on distributed regression for G-dWOLS with the inverse probability of treatment weight. The first data pooling strategy used a pool size of 3 across the different sites. The second data pooling strategy used a pool size of three for all sites. The third data pooling strategy adjusted the pool size to the site in order to reduce the number of patients excluded compared to the second data pooling strategy. However, the number of pools was reduced, which implied that there was less information. The data pooling strategies are described in Table 2 of the main document. The results of the different methods were compared to the results obtained with the gold-standard method using the original individual-level data.

In this additional simulation study, the estimand of interest is the optimal treatment, for which the vectors of the blip function parameters $\psi_1 = (\psi_{01}, \psi_{11})$ and $\psi_2 = (\psi_{02}, \psi_{12})$ are underlying. The blip function having the following expression $\gamma(x^{sim}, a^{sim}; \psi) = a^{sim}(\psi_{01} + \psi_{11}x^{sim} + \psi_{02}a^{sim} + \psi_{12}a^{sim}x^{sim})$ (second part of equation (S1)), the optimal treatment, which maximizes the blip function, can be estimated as:

$$\widehat{a^{sim}}_{opt} = -\frac{1}{2}\frac{(\psi_{01} + \psi_{11}x^{sim})}{(\psi_{02} + \psi_{12}x^{sim})} \qquad (S2)$$

### D. Results

The optimal treatment has been estimated for each one of the five methods. Figure S3 represents the average lowess (controlled by a smoother span) over the 300 independent simulated datasets, of the optimal treatment estimated by the five methods over the individual patients sorted by observed dose value. This figure shows that the distributed regression perform the best; then, the data pooling strategies 1 and 2 perform quite well, with a very low over-estimation, whereas the third data pooling strategy provide biased estimates, probably due to the difference of information.

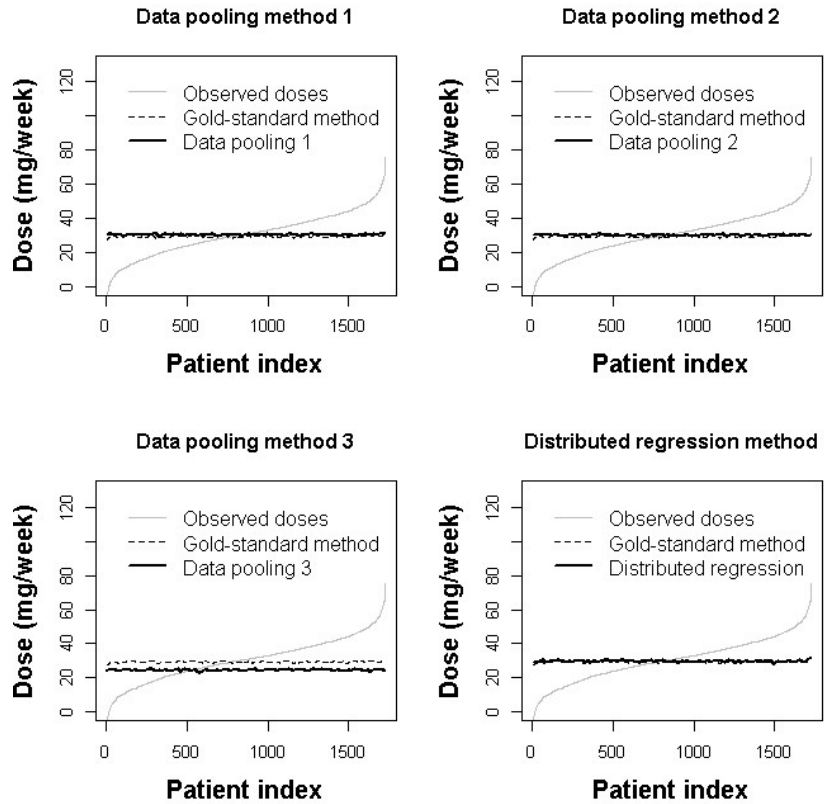

**Figure S3.** Additional simulation study based on the Warfarin example. Comparison of the optimal treatment estimated by the gold standard approach using individual data with no privacy preservation (dashed black line) and four privacy preserving methods (the three data pooling strategies and the distributed regression approach) over the patient index sorted by observed dose value. Note that all these estimates represent the mean optimal treatment over the 300 simulated datasets.